\documentclass[preprint,10pt]{elsarticle}
\usepackage[letterpaper,top=1.75cm,bottom=1.75cm,left=1.75cm,right=1.75cm,marginparwidth=1.75cm]{geometry}

\usepackage[utf8]{inputenc} 
\usepackage[T1]{fontenc}    
\usepackage{url}            
\usepackage{booktabs}       
\usepackage{amsfonts}       
\usepackage{nicefrac}       
\usepackage{microtype}      
\usepackage{lipsum}		
\usepackage{graphicx}
\usepackage{natbib}
\usepackage{doi}
\usepackage{physics}
\usepackage{xcolor}

\newcommand{\pd}[2]{\frac{\partial #1}{\partial #2}}

\usepackage{bm}

\journal{Elsevier}

\begin{document}

\begin{frontmatter}

\title{Physics-informed solution reconstruction in elasticity and heat transfer using the explicit constraint force method}

\author[1]{Conor Rowan}
\ead{conor.rowan@colorado.edu}
\author[1]{Kurt Maute}
\ead{kurt.maute@colorado.edu}
\author[1]{Alireza Doostan\corref{mycorrespondingauthor}}
\cortext[mycorrespondingauthor]{Corresponding author}
\ead{alireza.doostan@colorado.edu}
\affiliation[1]{address={Ann and H.J. Smead Department of Aerospace Engineering Sciences, University of Colorado, Boulder}}

\begin{abstract}
    
    Using neural networks to solve systems governed by ordinary and partial differential equations was first explored more than two decades ago, but has gained traction more recently under the banner of ``physics-informed neural networks'' (PINNs). One use case of PINNs is solution reconstruction, which aims to estimate the full-field state of a physical system from sparse measurements. Parameterized governing equations of the system are used in tandem with the measurements to regularize the regression problem. However, in real-world solution reconstruction problems, the parameterized governing equation may be inconsistent with the physical phenomena that give rise to the measurement data. In this work, we show that due to assuming consistency between the true and parameterized physics, PINNs-based approaches may fail to satisfy three basic criteria of interpretability, robustness, and data consistency. As we argue, these criteria ensure that (i) the quality of the reconstruction can be assessed, (ii) the reconstruction does not depend strongly on the choice of physics loss, and (iii) that in certain situations, the physics parameters can be uniquely recovered. In the context of elasticity and heat transfer, we demonstrate how standard formulations of the physics loss and techniques for constraining the solution to respect the measurement data lead to different ``constraint forces"---which we define as additional source terms arising from the constraints---and that these constraint forces can significantly influence the reconstructed solution. To avoid the potentially substantial influence of the choice of physics loss and method of constraint enforcement on the reconstructed solution, we propose the ``explicit constraint force method'' (ECFM) to gain control of the source term introduced by the constraint. We then show that by satisfying the criteria of interpretability, robustness, and data consistency, this approach leads to more predictable and customizable reconstructions from noisy measurement data, even when the parameterization of the missing physics is inconsistent with the measured system.
    \end{abstract}

\begin{keyword}
Solution reconstruction \sep Data assimilation \sep Constrained optimization \sep Physics-informed machine learning
\end{keyword}

\end{frontmatter}


%
%
\section{Introduction}

 One area of machine learning (ML) research with particular relevance to the engineering community is the use of neural networks (NNs) on physical systems governed by ordinary and partial differential equations. One can distinguish between approaches that solely rely on data and those that do not. Fully data-driven approaches make use of measurement or numerical simulation data relating inputs such as loading and/or boundary conditions to outputs such as displacement, velocity, and/or temperature. Instead of solving governing equations derived from conservation laws, the data is used to train a “surrogate model” relating inputs to outputs \cite{azizzadenesheli_neural_2024, bhaduri_stress_2022, lu_deeponet_2021, li_fourier_2021}. The alternative to these purely data-driven approaches is “physics-informed machine learning,'' in which physical laws in the form of differential equations are incorporated into the training process. In this setting, the neural network acts as a representation of the solution to a differential equation, and the parameters of the neural network are tuned according to some physics-based criteria. Here, the neural network replaces a more standard discretization (finite element, spectral, etc.) in the numerical solution. Note that in some cases, the physics-based loss may be supplemented by a loss incorporating measurement data. In general, data-driven approaches build a surrogate model that can solve the problem for a whole class of forcings and/or boundary conditions, though they sometimes make use of physical laws to this end \cite{wang_learning_2021}. In contrast, a physics-informed method finds a solution to a problem defined by specific forcings and/or boundary conditions by explicitly solving the governing equation. We will limit our attention to uses of neural networks that make use of physical laws in the service of solving particular initial/boundary value problems, as opposed to those that build surrogate models.

  Lagaris et al. performed prescient early studies of neural network discretizations of partial differential equations (PDEs) over two decades ago \cite{lagaris_artificial_1998, lagaris_neural-network_2000}. In their work, they discretize the PDE solution with a multilayer perceptron (MLP) neural network and determine its weights and biases by minimizing the sum of the squares of the PDE residual at discrete collocation points. Limited by available computational resources and difficulties with enforcing boundary conditions, they were restricted to simple physical models and geometries. More than 20 years later, the term ``physics-informed neural networks'' (PINNs) was coined when these methods were extended to more complex physical models and inverse problems \cite{raissi_physics-informed_2019}. This terminology is more widely adopted than the “Deep Galerkin method,” which employed similar methods and was introduced around the same time \cite{sirignano_dgm_2018}. In both cases, the solution of the PDE is taken as the minimum of the integral of the square of the strong form residual, with boundary/initial conditions enforced with penalties. These works laid the foundation for an explosion of interest in using ML techniques such as neural networks, automatic differentiation, and stochastic optimization to replace conventional numerical PDE solvers such as the finite volume, finite difference, and finite element methods. Follow-up works extended these methods to different physics, such as elasticity \cite{cai_deep_2020, kag_physics-informed_2024}, contact mechanics \cite{sahin_solving_2024}, heat transfer \cite{madir_physics_2024}, fluid mechanics \cite{jin_nsfnets_2021}, and homogenization \cite{leung_nh-pinn_2022}. Owing to concerns about the scalability of the vanilla PINNs approach, domain decomposition methods were introduced and explored in \cite{ameya_d_jagtap_extended_2020, hu_when_2022, jagtap_conservative_2020}. Various other modifications to the network architecture and loss function have been proposed to improve the accuracy and convergence rates of the learned solution \cite{anagnostopoulos_residual-based_2024, wang_respecting_2024, li_physical_2023, gao_physics-informed_2022, du_evolutional_2021}.
  
  The vanilla PINNs framework and its variants have had success in solving a wide range of problems, but these methods leave some fundamental assumptions unquestioned. For example, they all rely on minimizing a loss constructed with the PDE residual and enforcing boundary/initial conditions with some form of penalty. The residual loss can impose overly strict continuity requirements on the solution, and constraints enforced with penalties may not be satisfied accurately. Thus, other physics-informed machine learning approaches have made use of different formulations of the PDE loss and techniques for boundary condition enforcement \cite{sukumar_exact_2022, wang_exact_2023}. For many physical systems of engineering interest, the solution of the differential equation corresponds to a minimum of an energy functional, which defines a convenient objective for discretized solutions. This technique was first introduced with neural network discretizations of the solution as the “Deep Ritz Method” \cite{e_deep_2017}. Here, the solution is discretized with a neural network, the energy is constructed as a function of the neural network parameters, and, using gradient descent, the NN parameters are found such that the energy is minimal. The Deep Ritz Method has been studied in the context of linear elasticity \cite{liu_deep_2023}, hyperelasticity \cite{abueidda_deep_2022, nguyen-thanh_deep_2020}, thermoelasticity \cite{lin_investigating_2024}, and fracture mechanics \cite{manav_phase-field_2024, ghaffari_motlagh_deep_2023}. One advantage of this method is that energy functionals always have lower-order spatial derivatives on the solution than the strong form of the governing equation, thus evaluations of the loss function and its gradient are cheaper when using automatic differentiation. 

  In addition to the strong form residual and energy loss functions, the weak form of the PDE can also be deployed with neural network discretizations. Instead of a scalar loss function, the weak form is a system of equations expressing the orthogonality of the PDE residual to a chosen basis. Like the energy formulation, the weak form benefits from lower-order spatial differentiation. Unlike the energy, however, the weak form exists for any differential equation. With variational physics-informed neural networks (VPINNs), the spacetime solution is parameterized as a neural network, the strong form is integrated against a suitable basis of test functions, and the parameters are found by minimizing the norm of the weak form residual \cite{kharazmi_variational_2019}. Follow-up works, such as \cite{kharazmi_hp-vpinns_2021, shang_deep_2022, shang_randomized_2023, khodayi-mehr_varnet_2019}, explore the Petrov-Galerkin weak form loss in the context of a variety of other problems from engineering mechanics. Inspired by adversarial training from computer vision, an interesting alternative formulation is \cite{zang_weak_2020}, where the test function is a neural network trained to maximize the residual. By lowering the order of differentiation on the solution---and thus reducing continuity requirements---the weak form expedites training and expands the class of admissible solutions \cite{khodayi-mehr_varnet_2019}. 

  As the original PINNs paper showed \cite{raissi_physics-informed_2019}, it is straightforward to incorporate data into the loss function with an additional penalty term. By treating parameters of the physical model as trainable, this incorporation of the data loss gives rise to a powerful technique for solving ``inverse problems,'' which estimate unknown parameters of the model from measurement data. In this framework, the parameters can be estimated by simultaneously minimizing the PDE and data loss. The efficacy of this method was first demonstrated for estimating the unknown viscosity of the Navier-Stokes equation \cite{raissi_physics-informed_2019}. It has since between extended to conjugate heat transfer \cite{cai_physics-informed_2021}, contact mechanics \cite{sahin_solving_2024}, and beam bending \cite{zhou_data-guided_2024}.

  Whereas an inverse problem fits the solution of a PDE in the process of estimating the missing model parameters, a related problem called ``solution reconstruction'' does the opposite. With solution reconstruction, the goal is to obtain a full-field estimate of the state from sparse measurement data, which is frequently accomplished by estimating the model parameters. In other words, inverse problems seek the missing parameters and use the solution field en route, whereas solution reconstruction uses the parameters to predict the solution. These two problems are equivalent when the model which generated the measurement data can be exactly recovered by the inverse problem. However, when the parameterized model is inconsistent with the measurement data, the recovered parameters may cease to be meaningful and are only a means to an end. A number of solution reconstruction problems have been explored in the PINNs community, but primarily under the assumption that the model which generated the data can be exactly recovered. In \cite{raissi_hidden_2020}, the velocity and pressure fields are interpolated with a neural network and estimated from image data. In \cite{yeung_physics-informed_2022}, neural networks are used to infer the full-field state of the hydraulic head of groundwater flow by estimating a spatially varying constitutive parameter. Similarly, \cite{ehlers_data_2024} estimates coefficients in the governing equation as a step in reconstructing the full-field state of the velocity field of shallow water waves. There are many other examples of using neural networks to interpolate measurement data for fluid systems that adopt similar methods (cf. \cite{rui_reconstruction_2023, hosseini_flow_2024, saldern_mean_2022, angriman_assimilation_2023, sliwinski_mean_2023, ohashi_multiple_2024}). An alternative approach taken in \cite{zhang_operator_2025} uses operator learning to build a reconstruction map that interpolates measurement data. An exception to the assumption of consistency between the true and parameterized model is found in \cite{zou_correcting_2024}, where a neural network is used to parameterize missing physics terms. We will return to a discussion of this work when more details of our proposed methodology are laid out. 

  The examples of solution reconstruction thus far all come from the fluids community. In solid mechanics, however, solution reconstruction techniques are of interest for structural health monitoring \cite{xiang_structural_2018, kralovec_review_2020, wu_data_2020} and digital twins \cite{tao_digital_2019, wang_digital_2023, ferrari_digital_2024}. In this setting, numerical simulations are often supplemented with data because there is some uncertainty as to the physics governing the real structure. Thus, a challenge inherent to solution reconstruction in this setting is developing full-field state reconstructions that are robust to imprecise parameterizations of the true underlying model. With PINNs-based methods for solution reconstruction, it is typically assumed that the parameterized model is consistent with the measurement data, but this may be an unrealistic assumption in the case of complex physics. There have been some attempts to handle misspecified physics outside the PINN's literature \cite{masud_physics-constrained_2023}, but only with variational problems. Apart from \cite{zou_correcting_2024}, dealing with genuine conflicts between the parameterized governing equation and measurement data is a problem that has received comparatively little treatment when using neural networks for solution reconstruction.

  In this work, we investigate the solution reconstruction problem for static systems from solid mechanics and heat transfer. We propose three criteria that any effective solution reconstruction technique should satisfy---namely, interpretability, robustness, and data consistency--- and then show that standard methods from the physics-informed machine learning literature may fail to satisfy these criteria. These issues are especially pronounced when the system that generates the measurement data is not accurately modeled or parameterized in the reconstruction problem. Our primary contributions are:

\begin{enumerate}
    \item We show that the variational form of a PDE cannot be used as a loss function in the PINNs-based solution reconstruction problem;
    \item We demonstrate that constraints enforced with penalty methods and Lagrange multipliers introduce source terms to the system, and that the form of these source terms depends on numerical techniques for solving the solution reconstruction problem;
    \item We show that the idea of ``constraint forces'' can be used to explain the sensitivity of the reconstructed solution to the choice of physics loss and method of constraint enforcement;
    \item We propose that reconstructed solutions should obey a ``minimum constraint force'' principle and introduce a new method for solution reconstruction based on minimizing the constraint force; 
    \item By controlling the form of the source terms introduced by the constraint, this ``explicit constraint force method'' (ECFM) is shown to satisfy the three desiderata we lay out for a solution reconstruction method.
\end{enumerate}

  The remainder of the paper is organized as follows. In Section 2, we sketch the mathematical details of the abstract solution reconstruction problem, then propose three desiderata for a solution methodology. In Section 3, we review common neural network-based approaches to solving this problem from the literature and discuss different formulations of the physics loss and techniques for enforcing constraints. In Section 4, we show some of the failure modes of these standard approaches on a simple model problem from linear elasticity. In Section 5, we identify the causes of these failures. In Section 6, a novel approach to the solution reconstruction problem is proposed, which remedies the aforementioned shortcomings and satisfies the three criteria laid out previously. In Section 7, we demonstrate the method on three problems from linear and nonlinear problems in elasticity and heat transfer.




\section{Problem setup}

  The objective of solution reconstruction is to infer the full-field state of a system governed by an ordinary or partial differential equation from sparse measurements. We aim to incorporate knowledge of this underlying differential equation to reconstruct the solution from measurement data. In this work, the physical systems from which measurements are taken are supposed to obey a time-independent boundary value problem (BVP) of the form

\begin{equation} \label{bvp}
\begin{aligned}
   & \mathcal{N} \Big( \mathbf{u}
    (\mathbf{x}) \Big) + \mathbf{s}(\mathbf{x}) = \mathbf{0}, \quad \mathbf{x} \in \Omega,  \\
   & \mathbf{u}(\mathbf{x}) = \mathbf{g}(\mathbf{x}), \quad \mathbf{x} \in \partial \Omega,
\end{aligned}
\end{equation}

\noindent where $\mathcal{N}$ is a linear or nonlinear differential operator, $\mathbf{u}(\mathbf{x}) \in \Omega \subset \mathbb{R}^D$ is the state$, \mathbf{s}(\mathbf{x})$ is a source term, $\mathbf{g}(\mathbf{x})$ specifies the Dirichlet boundary condition, and $\partial \Omega$ is the boundary of the domain $\Omega$. Though the BVP of Eq. \eqref{bvp} generates the measurement data, the mathematical form of the differential operator, source term, and/or boundary condition may all be unknown or partially known. If this were not the case, the differential equation could be directly solved to infer the full-field state, rendering solution reconstruction unnecessary. In this work, we will assume that the system has known Dirichlet boundaries but an unknown differential operator $\mathcal{N}$ and source term $\mathbf{s}(\mathbf{x})$. The assumption of Dirichlet boundaries facilitates analysis but does not meaningfully limit our findings or proposed method. We note that this assumption implies the geometry of the domain $\Omega$ is known.

  We assume that the state of the system is measured at points $\{\mathbf{x}_i\}_{i=1}^C$ with $\mathbf{x}_i\in \Omega$. In general, these measurements will be noisy and are given by 

\begin{equation} \label{measurement}
    \mathbf{ v}_i = \mathbf{u}(\mathbf{x}_i) + \boldsymbol{\xi}_i, \quad \boldsymbol \xi_i \overset{\text{i.i.d.}}{\sim} \mathcal{D}.
\end{equation}

The measurement noise at each point ($\boldsymbol \xi_i$) is independently distributed according to an arbitrary symmetric distribution $\mathcal{D}$ with finite variance $\boldsymbol \sigma$. The system state $\mathbf{u}(\mathbf{x})$ obeys the unknown boundary value problem given in Eq. \eqref{bvp}. The measurements are assumed to be sparse, meaning that a method relying only on the data (such as polynomial regression) may yield an untrustworthy reconstruction of the full-field state. To regularize the regression problem, we require \textit{some} knowledge of the differential equation that describes the system state. In the context of engineering mechanics, conservation of mass, momentum, and energy typically govern the system dynamics but do not uniquely specify the differential operator $\mathcal{N}$. Different kinematic and constitutive models lead to a diverse set of operators, thus, uncertainty of this form is common. No physical laws govern the source term, so knowledge of its mathematical form must come from familiarity with the system or from the measurement data itself. We will assume throughout our study that the differential operator and source term are parameterized, though this parameterization may not be capable of exactly recovering the system in Eq. \eqref{bvp}. This means that our model for the BVP is specified up to a set of unknown constants. The solution reconstruction problem is then governed by 

\begin{equation} \label{assimilation}
\begin{aligned}
  &  \mathcal{G}\Big( \mathbf{w}(\mathbf{x}) ; \boldsymbol{\epsilon}_1 \Big) +  \mathbf{b}(\mathbf{x};\boldsymbol{\epsilon}_2) = \mathbf{0}, \quad \mathbf{x} \in \Omega, \\
  &  \mathbf{w}(\mathbf{x}) = \mathbf{g}({\mathbf{x}}), \quad \mathbf{x} \in \partial \Omega, \\
  &  \mathbf{ v}_i - \alpha \boldsymbol{\sigma} \leq \mathbf{w}(\mathbf{x}_i) \leq \mathbf{v}_i + \alpha \boldsymbol{\sigma}, \quad i=1,\dots,C,
\end{aligned}
\end{equation}

\noindent where the ``physics parameters'' $\boldsymbol{\epsilon} = [\boldsymbol{\epsilon}_1,\boldsymbol{\epsilon}_2]$ parameterize the differential operator $\mathcal{G}$ and source term $\mathbf{b}(\mathbf{x})$, and $\mathbf{v}_i$ are the noisy measurements taken from the real system. The parameter $\alpha$ in Eq. \eqref{assimilation} is chosen such that $P(| \mathbf{v}_i - \mathbf{u}(\mathbf{x}_i) | \geq \alpha \boldsymbol{\sigma} )$ is below a specified threshold. This simplifies analysis by transforming the stochastic optimization problem into a deterministic, inequality-constrained problem. This inequality constraint forces the reconstruction to satisfy the measurement while accounting for noise. The goal of the solution reconstruction problem is to find the solution $\mathbf{w}(\mathbf{x})$ and the physics parameters $\boldsymbol \epsilon$. We emphasize that the parameterized operator $\mathcal{G}\Big( \mathbf{w}(\mathbf{x});\boldsymbol{\epsilon_1}\Big)$ and source term $\mathbf{b}(\mathbf{x};\boldsymbol{\epsilon}_2)$ represent the analyst's best guess as to the physics that generates the measurement data. For example, consider the governing equation for a linearly elastic bar with variable modulus:

\begin{equation}\label{right}
    \pd{}{x}\qty( (1+x)\pd{u}{x} ) + x\sin(\pi x) =0, \quad u(0)=u(1)=0. 
\end{equation}

This is a specific instance of Eq. \eqref{bvp}. Say that the physical model used to reconstruct the solution from the measurement data in Eq. \eqref{assimilation} is parameterized as follows:

\begin{equation}\label{wrong}
    \epsilon_1 \frac{\partial^2 w}{\partial x^2} + \epsilon_2 \sin(\pi x) = 0,\quad u(0)=u(1)=0.
\end{equation}

We cannot recover the true BVP of Eq. \eqref{right} for any choice of the parameters $\boldsymbol{\epsilon}$ in Eq. \eqref{wrong} because the parameterizations of the operator and source term are inaccurate. The discrepancy comes from missing a multiplicative factor in the source term and the assumption of constant material. More extreme discrepancies in the operator arise from using different physics altogether, such as a linearly elastic parameterized BVP when the data was generated from a hyperelastic model. We believe potential conflicts between the parameterized model and the true physical system are the norm, not an exception.

  Note that $\mathbf{u}(\mathbf{x})$ is the true but unknown state which generates measurements, and $\mathbf{w}(\mathbf{x})$ is the reconstruction which interpolates the sparse measurements with the help of the parameterized physics. In the ideal case, we want the recovered solution to match the true solution, i.e., $\mathbf{w}(\mathbf{x})=\mathbf{u}(\mathbf{x})$. When there is no choice of parameters $\boldsymbol{\epsilon}$ such that Eq. \eqref{assimilation} recovers the BVP of Eq. \eqref{bvp}, this will not be possible. In this case, the goal of the solution reconstruction process is to obtain an optimal approximation of $\mathbf{u}(\mathbf{x})$ given the parameterized physics which govern $\mathbf{w}(\mathbf{x})$. Different notions of optimality will be discussed below.

  In general, it is not possible to obtain a closed-form solution to the problem given by Eq. \eqref{assimilation}. As we will show, there are a variety of mathematical and numerical techniques to obtain an approximate solution. To assess the efficacy of different approaches, we outline three basic criteria that any approach ought to respect:

\begin{enumerate}
    \item \textit{Interpretability:} There should be a clear and physically interpretable indicator of the quality of the reconstructed solution.
    \item \textit{Robustness to numerical solution:} The reconstructed solution should not strongly depend on the mathematical or numerical methods used to obtain an approximate solution of Eq. \eqref{assimilation}. In particular, the formulation of the solution to the differential equation and the technique of constraint enforcement should have an insignificant influence on the reconstructed solution.
    \item \textit{Data consistency:} As we will discuss, when the parameterized physics is inconsistent with the measurement data, a ``constraint force'' is introduced into the governing equation. The number of parameters discretizing the constraint force should match the number of measurements.

\end{enumerate}

  The first criterion states that the solution reconstruction method should provide the analyst with feedback as to the accuracy of the parameterization of the underlying BVP. In particular, we need to distinguish whether the parameterization in Eq. \eqref{assimilation} is consistent with the data generated from the system of Eq. \eqref{bvp}. If it is not consistent, we would like a measure of how far off it is. A quality measure of this sort should be given in units which are physically meaningful to the system under study (force, heat flux, etc.) to facilitate interpretation. This quality measure provides the analyst with feedback as to how trustworthy the reconstructed solution is. The second criterion states that the reconstructed solution should not strongly depend on mathematical aspects of the numerical solution. For example, the reconstruction from the data should be insensitive to whether the strong or weak form solution of the differential equation is used, or to whether constraints are enforced with Lagrange multipliers or penalty methods. Finally, the third criterion states that the form of the source term applied to the system to enforce constraints should be matched to the amount of measurement data. In other words, we seek to avoid overfitting the measurement data with needlessly complex constraint forces. We will show that satisfying this condition leads to unique solutions and identifiable physics parameters in certain idealized settings.

  Given the flexibility of neural networks as regression models, the ease of computing spatial derivatives with automatic differentiation, and the convenience of formulating Eq. \eqref{assimilation} as an optimization problem, the physics-informed machine learning community has taken an interest in solution reconstruction problems. It can be shown, however, that standard approaches to solving this problem in the PINNs literature fail to reliably satisfy any of the three criteria laid out above. These issues are especially pronounced when there are ``unknown unknowns'' in the physical model of Eq. \eqref{bvp}, meaning that the true underlying equation cannot be recovered by the parameterized physics. In this case, standard PINNs-based approaches may be ill-suited for solution reconstruction. To understand why this is the case, we must first review the most popular approaches taken in the physics-informed machine learning community. 

\begin{figure}[hbt!]
\centering
\includegraphics[trim = 50mm 65mm 40mm 40mm, clip, width=0.65\textwidth]{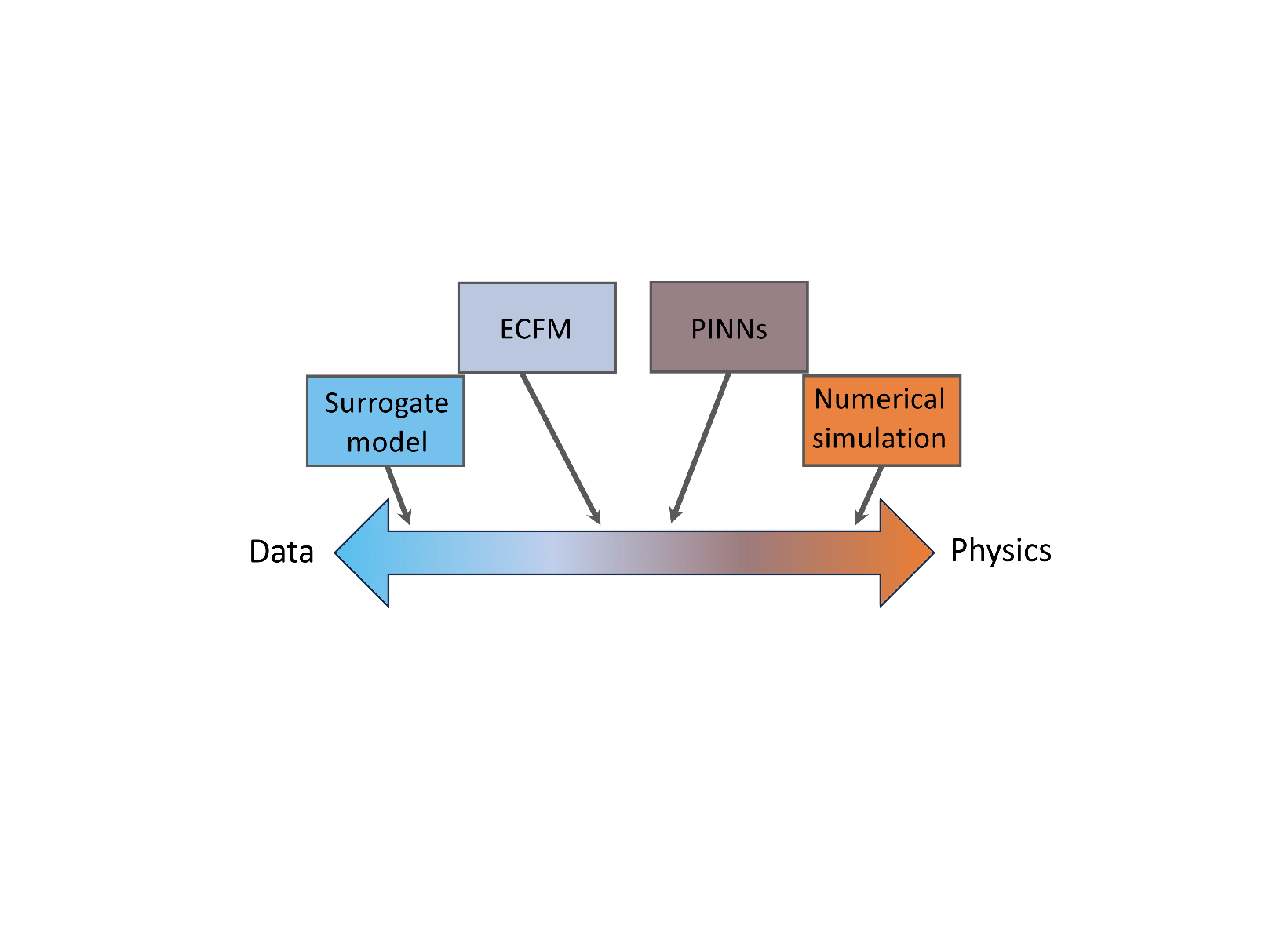}
\caption{Different techniques rely on different amounts of data and knowledge of the system's physics. Our explicit constraint force method (ECFM) does not rely on the assumption that the parameterized model is consistent with measurement data.}
\label{pp0}
\end{figure}

  As a final note, we see that the possibility of ``unknown unknowns'' in the physical model forces us to rely more heavily on the data than on the physics. This problem can be positioned in relation to other methods as shown in Figure \ref{pp0}. In taking the parameterized model to be consistent with the true physical system, PINNs assumes more knowledge of the physics than we do.

\section{Review of methods}

\subsection{Penalty methods}

  The most common ML-based approach to solution reconstruction is to transform Eq. \eqref{assimilation} into an unconstrained optimization problem with a penalty formulation. Assume that the solution $\mathbf{w}(\mathbf{x})$ is discretized in terms of a finite set of parameters $\boldsymbol{\theta}$. The details of the discretization are unimportant at this point---suffice it to say it could be a finite element, spectral, or neural network discretization (in the case of PINNs). An approximate solution, denoted by $\mathbf{\hat w}(\mathbf{x};\boldsymbol{\theta})$, is obtained through this discretization and a particular formulation of the solution to the boundary value problem. See Table \ref{tab:one} for a summary of some of the notation employed in this paper. Using the most common form of the penalty method \cite{raissi_physics-informed_2019}, the approximate solution is obtained with

\begin{equation} \label{penalty1}
\begin{split}
    \underset{\boldsymbol{\theta}, \boldsymbol{\epsilon}}{\text{argmin }} \bigg[ 
    & \frac{1}{2} \int \Big \lVert \mathcal{G}\Big( \mathbf{\hat w}(\mathbf{x};\boldsymbol{\theta}) ; \boldsymbol{\epsilon}_1 \Big) +  \mathbf{b}(\mathbf{x};\boldsymbol{\epsilon}_2) \Big \rVert^2 d\Omega \\
    & + \frac{\lambda_b}{2}\int_{\partial \Omega}\Big \lVert \mathbf{\hat w}(\mathbf{x};\boldsymbol{\theta}) - \mathbf{g}(\mathbf{x}) \Big \rVert^2 dS 
    + \frac{\lambda_d}{2} \sum_{i=1}^C\Big \lVert \mathbf{\hat w}(\mathbf{x}_i;\boldsymbol \theta) - \mathbf{v}_i \Big \rVert^2 \bigg].
\end{split}
\end{equation}


The two penalty parameters $\lambda_b$ and $\lambda_d$ weigh the contributions of the Dirichlet boundary condition error and data error to the total loss. Here, the parameterized differential equation in Eq. \eqref{assimilation} is solved by minimizing the integral of its squared error. We will call the first term in Eq. \eqref{penalty1} the ``strong form loss.'' Note that a common modification of this formulation is to build the Dirichlet boundaries into the discretization with

\begin{equation}\label{dirichlet}
\begin{aligned}
   & \mathbf{\hat w}(\mathbf{x};\boldsymbol{\theta}) = \mathbf{D}(\mathbf{x}) \mathbf{\hat W}( \mathbf{x};\boldsymbol{\theta}) + \mathbf{G}(\mathbf{x}), \\
   & \mathbf{D}(\mathbf{x}) = \mathbf{0}, \quad \mathbf{x} \in \partial \Omega ,\\
   & \mathbf{G}(\mathbf{x}) = \mathbf{g}(\mathbf{x}), \quad \mathbf{x} \in \partial \Omega.     
\end{aligned}
\end{equation}

The function $\mathbf{\hat W}( \mathbf{x};\boldsymbol{\theta})$ is a discretization of the solution which need not respect the Dirichlet boundary conditions. Constructing the functions $\mathbf{D}(\mathbf{x})$ and $\mathbf{G}(\mathbf{x})$ may be non-trivial on complex spatial domains, but strict enforcement of Dirichlet boundaries leads to improved accuracy of the solution \cite{wang_exact_2023, sukumar_exact_2022, sheng_pfnn_2021, berrone_enforcing_2023}. Moving forward, we will assume that the Dirichlet boundary conditions are built into the discretization per Eq. \eqref{dirichlet}. In this case, the loss and optimization problem can be written as

\begin{equation} \label{L1}
\begin{aligned}
      &  \underset{\boldsymbol{\theta},\boldsymbol{\epsilon}}{\text{argmin } } \mathcal{L}_1, \\
      &  \mathcal{L}_1 = \frac{1}{2} \int \Big \Vert \mathcal{G}\Big( \mathbf{\hat w}(\mathbf{x};\boldsymbol{\theta}) ; \boldsymbol{\epsilon}_1 \Big) +  \mathbf{b}(\mathbf{x};\boldsymbol{\epsilon}_2) \Big \rVert^2 d\Omega  + \frac{\lambda_d}{2} \sum_{i=1}^C\Big \lVert \mathbf{\hat w}(\mathbf{x}_i;\boldsymbol{\theta}) - \mathbf{v}_i \Big \rVert^2 .    
\end{aligned}
\end{equation}

Note that the penalty method enforces equality constraints on the measurements. It is possible to better handle the inequality constraints of Eq. \eqref{assimilation} with a penalty method by passing the data error into a monotonic function, which is zero until the magnitude of the error exceeds the noise threshold. This ensures that no penalty is incurred when the inequality constraint is respected \cite{sahin_solving_2024}. 

  In the ideal scenario of the problem given in Eq. \eqref{L1}, $\mathcal{G}(\cdot) = \mathcal{N}(\cdot)$ and $\mathbf{b}(\mathbf{x})=\mathbf{s}(\mathbf{x})$ so that at the solution, there is no conflict between satisfying the governing equation and the data constraint. Here, solution reconstruction and inverse problems are equivalent. However, in many realistic solution reconstruction problems, the true physical model is not known. The objective is to obtain a full-field state reconstruction from the measurement data, and the parameters only aid in this task. Given that the parameterized physics might be inaccurate, the parameters may or may not be physically meaningful. In Section 4, we will show that problems arise if one applies this methodology to situations where the true and parameterized models are inconsistent. In this setting, it is not possible to satisfy the physics and the data constraint simultaneously. Relaxing the assumption of consistency between the true and parameterized models is our point of departure from traditional problems in the PINNs literature, which we believe to be a more realistic reflection of real-world solution reconstruction problems. We require that a solution reconstruction method should be identifiable, interpretable, and robust even in this setting.

\begin{table}[h]
  \centering
  \begin{tabular}{|c|c|}
    \hline
    \textbf{Symbol} & \textbf{Meaning} \\
    \hline
    $\Omega$ & Computational domain \\
    $\mathbf{x}$ & Spatial coordinate \\
    $\mathbf{u}$ & True solution \\
    $\mathbf{s}$ & True source term \\
    $\mathcal{N}$ & True operator \\
    $\mathbf{v}_i$ & Measurement from true solution \\
    $ C $ & Number of measurements from true solution \\
    $\boldsymbol{\epsilon}$ & Physics parameters of reconstruction problem \\
    $\mathbf{w}$ & Reconstructed solution \\
    $\boldsymbol{\theta}$ & Discretization parameters for reconstructed solution \\
    $\mathbf{b}$ & Parameterized source term for reconstruction\\
    $\mathcal{G}$ & Parameterized operator for reconstruction \\
    $\mathbf{\hat w}$ & Approximation of reconstructed solution \\ 
    \hline
  \end{tabular}
  \caption{List of symbols used in the description of the true physical system, the reconstructed solution, and the numerical approximation of the reconstructed solution.}
  \label{tab:one}
\end{table}

\subsection{Hard constraint enforcement}

  Penalty methods rely on the penalty hyperparameter, which can lead to poorly-conditioned optimization problems if not chosen carefully \cite{krishnapriyan_characterizing_2021}. An alternative approach is to enforce the data (or physics) constraints exactly by solving a constrained optimization problem with Lagrange multipliers \cite{norman_constrained_2025}. The constrained optimization problem with the strong form loss is

\begin{equation} \label{constrained1}
\begin{aligned}
    &\underset{\boldsymbol{\theta},\boldsymbol{\epsilon}}{\text{argmin }} \frac{1}{2} \int \Big \Vert \mathcal{G}\Big( \mathbf{\hat w}( \mathbf{x};\boldsymbol{\theta}) ; \boldsymbol{\epsilon}_1 \Big) +  \mathbf{b}(\mathbf{x};\boldsymbol{\epsilon}_2) \Big \rVert^2 d\Omega \\ 
    &\text{s.t. } \mathbf{\hat w}(\mathbf{x}_i;\boldsymbol{\theta}) -  \mathbf{v}_i = 0, \quad i=1,\dots,C,
\end{aligned}
\end{equation}

\noindent where equality constraints are temporarily assumed for the sake of exposition. Note that equality constraints on the measurements imply there is no noise. By introducing the Lagrange multipliers $\{\boldsymbol{\lambda}_i\}_{i=1}^C$ and the Lagrange function $\mathcal{L}_2$, a solution to Eq. \eqref{constrained1} can be found with

\begin{equation} \label{L2}
\begin{aligned}
   & \mathcal{L}_2 = \frac{1}{2}\int \Big \lVert \mathcal{G}\Big( \mathbf{\hat w}(\mathbf{x};\boldsymbol{\theta}) ; \boldsymbol{\epsilon}_1 \Big) + \mathbf{b}(\mathbf{x};\boldsymbol{\epsilon}_2) \Big \rVert^2 d\Omega + \sum_{i=1}^C \boldsymbol{\lambda}_i \cdot \Big( \mathbf{\hat w}(\mathbf{x}_i;\boldsymbol{\theta}) - \mathbf{v}_i\Big), \\
   & \pd{\mathcal{L}_2}{\boldsymbol{\theta}} = \mathbf{0}, \quad \pd{\mathcal{L}_2}{\boldsymbol{\epsilon}} = \mathbf{0}, \quad \pd{\mathcal{L}_2}{\boldsymbol{\lambda}} = \mathbf{0}.
\end{aligned}
\end{equation}

Returning to the inequality-constrained problem corresponding to noisy measurements of the system, the optimization problem is 

\begin{equation} \label{constrained}
\begin{aligned}
    &\underset{\boldsymbol{\theta},\boldsymbol{\epsilon}}{\text{argmin }} \frac{1}{2} \int \Big \lVert \mathcal{G}\Big( \mathbf{\hat w}(\mathbf{x};\boldsymbol{\theta}) ; \boldsymbol{\epsilon_1} \Big) +  \mathbf{b}(\mathbf{x};\boldsymbol{\epsilon}_2) \Big \rVert^2 d\Omega \\ 
    &\text{s.t. } \mathbf{h}_i(\boldsymbol{\theta}) \leq \mathbf{0} \quad i=1,\dots,2C,
\end{aligned}
\end{equation}

\noindent where the inequality constraints $ \mathbf{v}_i - \alpha \boldsymbol{\sigma} \leq \mathbf{\hat w}(\mathbf{x}_i;\boldsymbol{\theta}) \leq \mathbf{v}_i + \alpha\boldsymbol{\sigma} $ are simple to rearrange and write in this standard form. This is accomplished with

\begin{equation} \label{constraints}
    \mathbf{h}_i(\boldsymbol{\theta}) = \begin{cases} \mathbf{\hat w}(\mathbf{x}_i,\boldsymbol{\theta}) - \mathbf{v}_i - \alpha \boldsymbol{\sigma}, & i=1,\dots,C ,\\
    \mathbf{v}_{i-C} - \alpha \boldsymbol{\sigma} - \mathbf{\hat w}(\mathbf{x}_{i-C},\boldsymbol{\theta}), &  i=C+1,\dots,2C. 
    \end{cases}
\end{equation}

The optimization problem with inequality constraints can be solved with the help of slack variables. The constraints are modified to read

\begin{equation} \label{constraints1}
    \mathbf{h}_i(\boldsymbol{\theta}) + \mathbf{s}_i^2 = \mathbf{0}, \quad i=1,\dots,2C,
\end{equation}

\noindent where $\{\mathbf{s}_i\}_{i=1}^{2C}$ are new optimization variables which transform the inequality constraints into equality constraints and their product is element-wise. Now, following a standard Lagrange multiplier approach to an equality-constrained problem, the solution must be a stationary point of the new objective function:

\begin{equation} \label{L3}
\begin{aligned}
    &\mathcal{L}_3(\boldsymbol{\theta},\boldsymbol{\epsilon},\boldsymbol{\lambda},\mathbf{s}) = \frac{1}{2} \int \Big \lVert \mathcal{G}\Big( \mathbf{\hat w}(\mathbf{x};\boldsymbol{\theta}) ; \boldsymbol{\epsilon}_1 \Big) +  \mathbf{b}(\mathbf{x};\boldsymbol{\epsilon}_2) \Big \rVert^2 d\Omega + \sum_{i=1}^{2C} \boldsymbol{\lambda}_i\cdot \Big( \mathbf{h}_i + \mathbf{s}_i^2\Big),\\
    &  \pd{\mathcal{L}_3}{\boldsymbol{\theta}} = \mathbf{0}, \quad  \pd{\mathcal{L}_3}{\boldsymbol{\epsilon}}  = \mathbf{0}, \quad   \pd{\mathcal{L}_3}{\boldsymbol{\lambda}}  = \mathbf{0}, \quad  \pd{\mathcal{L}_3}{\mathbf{s}} = \mathbf{0}.
\end{aligned}
\end{equation}

An additional requirement of a solution to the inequality-constrained problem is $\boldsymbol{\lambda} \geq \boldsymbol{0} $, which is called the ``dual feasibility'' condition \cite{ruszczynski_nonlinear_2006}. 

\paragraph{Remark} We note that in Eqs. \eqref{constrained1} and \eqref{constrained}, the data is treated as the constraint and the physics as an objective. We opt for this formulation given that the variance of the measurement noise is known, and the parameterization of the physics may be inexact. Thus, the data is treated as the ground truth, and the range of the inequality constraint is well-defined. If the objective and constraint were inverted, meaning that we minimize error with the measurement data subject to a physics constraint, we would be encouraging the reconstruction to exactly pass through the data despite the noise and enforcing a potentially inaccurate constraint with no clear way to set up inequality bounds.

  When the minimum of the objective does not naturally satisfy the constraints, a penalty method will not enforce constraints exactly. Despite this, the penalty parameter can be chosen to lead to small constraint violations. Conversely, a solution to the Lagrange multiplier systems of Eqs. \eqref{L2} or \eqref{L3} corresponds to exact satisfaction of the constraints. Thus, the primary advantage of the method of Lagrange multipliers is accuracy in constraint enforcement \cite{makridakis_deep_2024}. Other similar methods for constraint enforcement are SA-PINNs and the Augmented Lagrangian Method, which have been explored in \cite{mcclenny_self-adaptive_2023, son_enhanced_2023, basir_physics_2022}. Nitsche's method has also been investigated in the context of physics-informed machine learning \cite{liao_deep_2021}. These methods all exhibit certain similarities to penalty and Lagrange multiplier methods and have been shown to accurately enforce constraints in the context of neural network discretizations of PDE solutions.

\subsection{Beyond the strong form loss}

  In the methods discussed above, the solution of the differential equation is obtained by minimizing the integral of the strong form error. Minimizing the strong form loss is not the only way to obtain a solution, however. The weak form of the differential equation is constructed by requiring that the strong form residual is orthogonal to a given function space. An indexed set of functions $\{\mathbf{f}_j(\mathbf{x})\}_{j=1}^M$ forms a basis for the ``test space,'' and the condition that strong form error is orthogonal to this space reads

\begin{equation}\label{weak}
R_j = \int \mathcal{G}\Big( \mathbf{\hat w}(\mathbf{x};\boldsymbol{\theta}) ; \boldsymbol{\epsilon}_1 \Big) \cdot \mathbf{f}_j +  \mathbf{b}(\mathbf{x};\boldsymbol{\epsilon}_2) \cdot \mathbf{f}_j d\Omega = 0 \quad j=1,\dots,M.
\end{equation}

Eq. \eqref{weak} is the weak form residual system of the parameterized BVP. By transforming the solution of the weak form system into an optimization problem, another approach to approximate a solution to Eq. \eqref{assimilation} is given by

\begin{equation} \label{weakopt}
\begin{aligned}
    &\underset{ \boldsymbol{\theta} , \boldsymbol{\epsilon} }{\text{argmin }} \frac{1}{2} \sum_{j=1}^M \Big( \int \mathcal{G}\Big( \mathbf{\hat w}(\mathbf{x};\boldsymbol{\theta}) ; \boldsymbol{\epsilon}_1 \Big)\cdot \mathbf{f}_j +  \mathbf{b}(\mathbf{x};\boldsymbol{\epsilon}_2)\cdot \mathbf{f}_j d\Omega \Big)^2 \\
    &\text{s.t. } \mathbf{h}_i(\boldsymbol{\theta}) \leq \mathbf{0} \quad i=1,\dots,2C.
\end{aligned}
\end{equation}

The constraints can be enforced with any of the techniques shown above. For example, in the case of equality constraints, we can use Lagrange multipliers to find the solution as

\begin{equation} \label{L4}
    \begin{aligned}
    &\mathcal{L}_4 = \frac{1}{2} \sum_{j=1}^M \Big( \int \mathcal{G}\Big( \mathbf{\hat w}(\mathbf{x};\boldsymbol{\theta}) ; \boldsymbol{\epsilon}_1 \Big)\cdot \mathbf{f}_j +  \mathbf{b}(\mathbf{x};\boldsymbol{\epsilon}_2)\cdot \mathbf{f}_j d\Omega \Big)^2 + \sum_{i=1}^C \boldsymbol{\lambda}_i \cdot \Big( \mathbf{u}(\mathbf{x}_i; \boldsymbol{\theta}) - \mathbf{v}_i \Big),\\
    &\pd{\mathcal{L}_4}{\boldsymbol{\theta}} = \mathbf{0}, \quad \pd{\mathcal{L}_4}{\boldsymbol{\epsilon}}= \mathbf{0}, \quad   \pd{\mathcal{L}_4}{\boldsymbol{\lambda}} = \mathbf{0}.
    \end{aligned}
\end{equation}

This formulation is advantageous as it can often be used to lower the order of differentiation on the solution by transferring derivatives onto the test functions. This simultaneously decreases continuity requirements and speeds up computations involving automatic differentiation \cite{messenger_weak_2021}. We will call the objective in Eq. \eqref{weakopt} the ``weak form loss."

  There is a third notion of a solution to a BVP that will be of use to us. For a certain class of partial differential equations, there exists an ``energy density'' $\Psi\Big( \mathbf{w}(\mathbf{x}) \Big)$ which can be used to define the ``total potential energy'' of the system as

\begin{equation} \label{Pi}
    \Pi\Big( \mathbf{w}(\mathbf{x}) \Big) = \int \Psi\Big(\mathbf{w}(\mathbf{x})\Big) -  \mathbf{b}(\mathbf{x})\cdot \mathbf{w}(\mathbf{x}) d\Omega. 
\end{equation}

By the definition of the energy density and the total potential, the strong form of the governing equation can be recovered as a condition of stationarity of the energy functional

\begin{equation*} \label{variation}
    \delta \Pi = 0.
\end{equation*}

For many problems in elasticity and heat transfer, energy densities exist, and their stationary points are minima. This means that a solution to the differential equation can be obtained by directly minimizing the energy functional. Thus, in the case of the discretized solution and parameterized BVP, Eq. \eqref{Pi} becomes

\begin{equation}
    \hat \Pi(\boldsymbol{\theta},\boldsymbol{\epsilon}) = \int \Psi\Big(\mathbf{\hat w}(\mathbf{x};\boldsymbol{\theta}) ;\boldsymbol{\epsilon}_1\Big) -  \mathbf{b}(\mathbf{x};\boldsymbol{\epsilon}_2)\cdot \mathbf{\hat w}(\mathbf{x};\boldsymbol{\theta}) d\Omega. 
\end{equation}

A third notion of an approximate solution to the solution reconstruction problem is then

\begin{equation*} \label{energyopt}
\begin{aligned}
    &\underset{\boldsymbol{\theta},\boldsymbol{\epsilon}}{\text{argmin }} \hat \Pi \\
    &\text{s.t. } \mathbf{h}_i(\boldsymbol{\theta}) \leq \boldsymbol{0}.
\end{aligned}
\end{equation*}

We will call the objective here the ``energy loss.'' Once again, if the measurements are noiseless and the constraints are enforced with Lagrange multipliers, the solution to this problem is given by 

\begin{equation} \label{L5}
\begin{aligned}
   & \mathcal{L}_5 = \int \Psi\Big(\mathbf{\hat w}(\mathbf{x};\boldsymbol{\theta}) ;\boldsymbol{\epsilon}_1\Big) -  \mathbf{b}(\mathbf{x};\boldsymbol{\epsilon}_2)\cdot \mathbf{\hat w}(\mathbf{x};\boldsymbol{\theta}) d\Omega + \sum_{i=1}^C \boldsymbol{\lambda}_i \cdot \Big( \mathbf{u}(\mathbf{x}_i;\boldsymbol{\theta}) - \mathbf{v}_i\Big) ,\\
   & \pd{\mathcal{L}_5}{\boldsymbol{\theta}} = \mathbf{0}, \quad \pd{\mathcal{L}_5}{\boldsymbol{\epsilon}} = \mathbf{0}, \quad  \pd{\mathcal{L}_5}{\boldsymbol{\lambda}} = \mathbf{0}.
\end{aligned}
\end{equation}

The strong form, weak form, and energy loss are the three primary approaches to formulating a solution to a boundary value problem, and thus can all be used in the context of the parameterized system used to regularize the sparse regression problem of solution reconstruction. Penalty and Lagrange multiplier formulations will be taken as the two primary methods to enforce constraints, with methods like SA-PINNs, the Augmented Lagrangian, and Nitsche's Method falling somewhere ``in between'' these two approaches.


\section{Characterizing the three criteria}

  Previously, we claimed that the methods reviewed above fail to deliver the three desiderata of an effective solution reconstruction technique. To explore this claim, we will work with a simple one-dimensional model problem. The displacement $u(x)$ of a linearly elastic bar with two fixed ends and Young's Modulus $E(x)$ under the action of a body force $s(x)$ is governed by

\begin{equation} \label{model}
\pd{}{x}\qty(E(x) \pd{u}{x})+ s(x) = 0 ,\quad u(0) = u(1) = 0.
\end{equation} 

The domain here is $\Omega \in [0,1]$. When the modulus and source term are specified, this equation can be solved to generate the measurement data. To interpolate the measurement data, we assume a parameterized BVP of the form

\begin{equation} \label{modelproblem}
    \begin{aligned}
       & \frac{\partial^2 w}{\partial x^2} + b(x;\boldsymbol{\epsilon}) = 0, \quad x \in [0,1], \\
       & w(0)=w(1)=0 ,\\
       & w(x_i) - v_i = 0 ,\quad i=1,\dots,C,
    \end{aligned}
\end{equation}

\noindent where $\boldsymbol \epsilon$ are the physics parameters. For the following discussion, we choose to work only with a parameterized source term and equality constraints on the measurement data. The displacement is discretized with

\begin{equation} \label{fourier}
\hat w(x;\boldsymbol{\theta}) = \sum_{i=1}^N \theta_i \sin(i \pi x).
\end{equation}

The sine series discretization satisfies the homogeneous Dirichlet boundary conditions automatically. Note that the nature of the discretization has no influence on the validity of the following results, so we choose a particularly simple form.

\subsection{Interpretability}

  A reconstructed solution is interpretable if it provides a quality measure phrased in units inherent to the system under study. This measure should give insight into whether the parameterized physics is consistent with the underlying model that generated the data. If the parameterization is consistent with the data, the analyst can have more confidence in the reconstruction of the system state between measurement points. We argue that a measure of the magnitude of additional source terms required for the system to satisfy the constraints is one such interpretable measure. To explore whether PINNs-type methods offer some such measure, we assume that the spatial form of the body force in Eq. \eqref{modelproblem} is known, but its magnitude is uncertain. Using the penalty method given in Eq. \eqref{L1}, the objective and optimality conditions are 

\begin{equation} \label{L1p}
    \begin{aligned}
        & \mathcal{L}_1 = \frac{1}{2}\int_0^1\qty( \frac{\partial^2 \hat w}{\partial x^2} + \epsilon b(x)  )^2 dx + \frac{\lambda_d}{2}\sum_{i=1}^C\Big( \hat w(x_i;\theta) - v_i \Big)^2, \\ &
        \pd{\mathcal{L}_1}{\boldsymbol{\theta}} = \mathbf{0}, \quad  \pd{\mathcal{L}_1}{\boldsymbol{\epsilon}} = \mathbf{0}.
    \end{aligned}
\end{equation}

\begin{figure}[hbt!]
\centering
\includegraphics[width=0.98\textwidth]{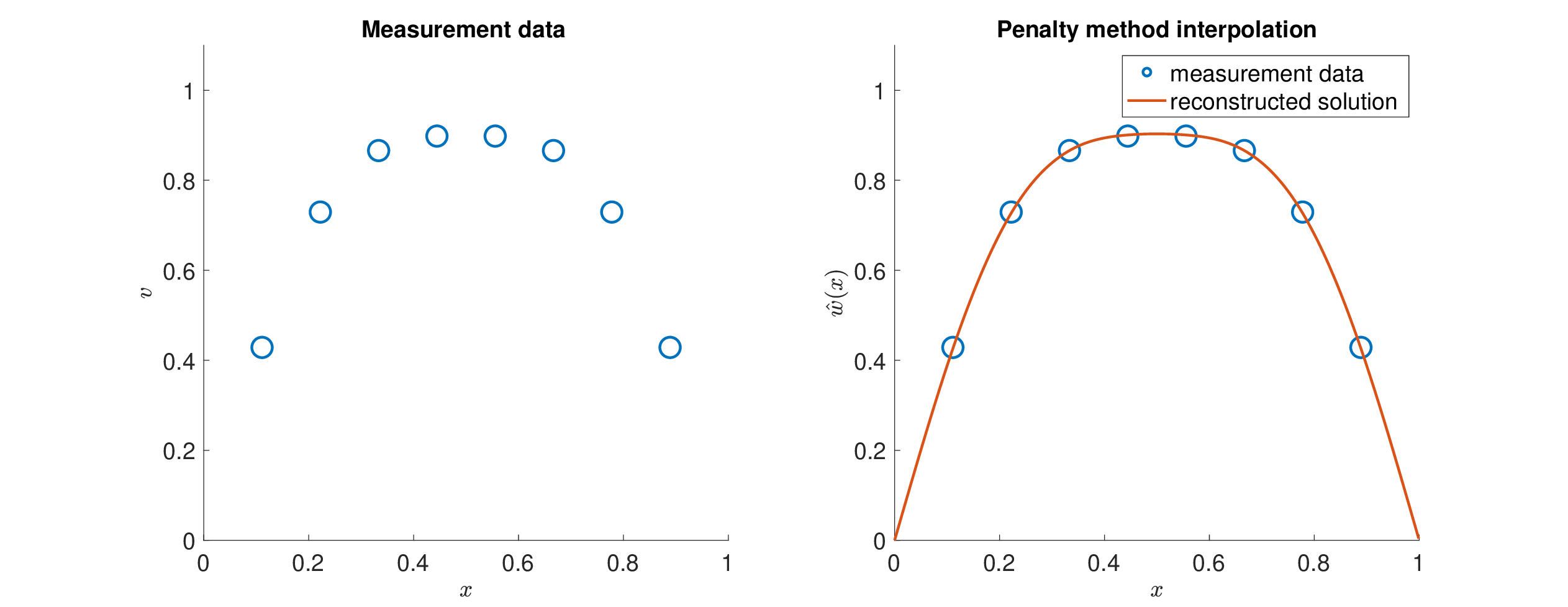}
\caption{Measurement data for the 1D linearly elastic BVP (left) and interpolation of the measurement data with the assumed body force model and penalty parameter (right). It is not clear whether the constraints are enforced as a result of recovering the physics parameters or from the penalty term.}
\label{pp2}
\end{figure}

To gauge the interpretability of reconstructions from the penalty method, we consider the following scenario. We take $C=8$ measurements from the unknown solution $u(x)$ without noise. This data is shown in Figure \ref{pp2}. Familiarity with the 1D linearly elastic BVP suggests that a reasonable guess for a body force that would produce this displacement is $b(x;\epsilon) = \epsilon \sin \pi x$. Using this parameterization of the source term, a penalty parameter of $\lambda_d=3\text{e}4$, and the discretization of the displacement in Eq. \eqref{fourier}, we can reconstruct the solution according to Eq. \eqref{L1p}. The results of this are shown in Figure \ref{pp2}. The recovered body force magnitude is $\epsilon = 9.87$, which suggests that a nonzero body force contributes to minimizing the strong form loss and, therefore, that the parameterization is not wholly inconsistent with the data. For example, if the optimizer found the solution $\epsilon=0$, this would suggest that satisfying the data constraint is entirely at odds with enforcing the parameterized physics. In addition to the non-zero body force, the reconstruction accurately satisfies the equality constraints on the measurement data. However, it is not clear whether the penalty term or the choice of the body force magnitude is responsible for the satisfaction of the constraints. One approach to answering this question would be to iteratively decrease the penalty parameter and observe if the reconstruction continues to satisfy the constraints. If the reconstructed solution still satisfies the constraints as the penalty parameter approaches zero, it is clear that the parameterization is consistent with the measurement data, as an accurate estimate of the source term gives rise to a model that reproduces the data. Note that this would be the case when assuming that the true model is recoverable. In general, however, the constraints will not be enforced through the source term alone, as the parameterization of the physics may not be rich enough to reproduce exactly the true system's operator and/or source term. Thus, the problem with this approach is that the penalty term is not a physical quantity, so it is not straightforward to establish guidelines as to what a ``large'' or ``small'' penalty parameter $\lambda_d$ is. If the penalty parameter in Eq. \eqref{L1p} could be interpreted as a force, heat flux, or some other physical quantity, it would provide insight into the extent of the missing physics in the parameterization of the system. But in the current form of Eq. \eqref{L1p}, no such interpretation is possible. This can be seen by noting that the units of the penalty parameter must be $\text{Force}^2/\text{Length}^3$ for Eq. \eqref{L1p} to be dimensionally consistent. Consequently, there is no quality measure given in meaningful physical units which allows us to understand the accuracy of the parameterization of the missing physics. Numerical experimentation could uncover the relationship between the penalty parameter and the constraint error, but the penalty parameter is not a physical quantity. Thus, even this would not provide an interpretable measure of the accuracy of the parameterization of the missing physics.

  It is simple to extend this argument to the case of constraint enforcement with Lagrange multipliers. The objective for the model problem with exact constraint enforcement is 

\begin{equation*} \label{misc}
    \begin{aligned}
       & \underset{\boldsymbol{\theta},\epsilon}{\text{argmin }} \frac{1}{2}\int_0^1 \qty( \frac{\partial^2 \hat w}{\partial x ^2} + \epsilon b(x) )^2dx \\
       & \text{s.t. } \hat w(x_i;\boldsymbol{\theta}) - v_i = 0, \quad i=1,\dots,C.
    \end{aligned}
\end{equation*}

When this constrained optimization problem is solved with a standard optimizer---which involves forming the Lagrange function given in Eq. \eqref{L2} and finding a stationary point with respect to the primal variables $\boldsymbol \theta$ and $\epsilon$ and the dual variables $\boldsymbol \lambda$---the results are even less interpretable. A nonzero $\epsilon$ indicates that the source term is consistent with the measurement data to some extent. If the optimizer zeroed the source term, it would suggest that the parameterization is so inaccurate that it drives up the strong form error. When the details of handling constraints are black-boxed by the optimization tool, the analyst is given minimal feedback as to the relative contributions of the physics parameter $\epsilon$ and the constraint terms in satisfying the constraints. At least for the penalty approach, it is possible to assess the relation between the penalty magnitude and constraint error, even if it is not clear what a large or small penalty is for the particular system under study. As such, the solution reconstructed with Lagrange multipliers provides no physical measure of the accuracy of the assumptions on the source term. By our definition, the reconstructed solutions from both the penalty method and Lagrange multipliers are not interpretable. Though this is demonstrated on a 1D example, these arguments generalize to 2D and 3D problems.

\subsection{Robustness}

  We define a reconstruction method to be robust when the solution does not exhibit strong sensitivity to the numerical methods used to solve the problem. In other words, whether the solution to the BVP is formulated using the strong form, weak form, or energy loss, the results should not differ significantly, and the character of the solution should not be strongly influenced by the choice of physics loss. Similarly, the solution should not exhibit strong sensitivity to whether constraints are enforced with a penalty method, Lagrange multipliers, or otherwise. To see the sensitivity of the reconstructed solution to the choice of physics loss, consider the following discretized energy function:

\begin{equation*} \label{energy1}
    \hat \Pi\Big( \hat w(x;\boldsymbol{\theta}) ; \epsilon \Big) = \int_0^1 \frac{1}{2}\qty(\pd{\hat w}{x})^2 - \epsilon b(x) \hat w(x;\boldsymbol{\theta}) dx.
\end{equation*}

This is the variational energy corresponding to the 1D linearly elastic model problem with homogeneous material and parameterized body force. Using this discretized energy with Lagrange multipliers to enforce equality constraints, the solution reconstruction problem reads

\begin{equation} \label{L5p}
\begin{aligned}
      &  \mathcal{L}_5 =  \int_0^1 \frac{1}{2}\qty(\pd{\hat w}{x})^2 - \epsilon b(x) \hat w(x;\boldsymbol{\theta}) dx + \sum_{i=1}^C \lambda_i\Big( \hat w(x_i;\boldsymbol{\theta}) - v_i \Big), \\
      &  \pd{\mathcal{L}_5}{\boldsymbol{\theta}} = \mathbf{0}, \quad  \pd{\mathcal{L}_5}{\boldsymbol{\epsilon}} = \mathbf{0}.
\end{aligned}
\end{equation}

Solving this minimization problem, which appears to be a simple reformulation of Eq. \eqref{L2p}, presents issues. Unlike the strong form loss, the energy in Eq. \eqref{L5p} is unbounded when minimized over $\epsilon$. With a source term specified by the parameter $\epsilon$, a solution to a differential equation can be obtained by minimizing its corresponding variational energy over the parameters $\boldsymbol{\theta}$. But unlike the strong form loss, choosing $\epsilon$ to lower the energy does not correspond to a ``better'' solution. This is an interesting and subtle difference between the different formulations of the BVP that we will explore in more detail.

\begin{figure}[hbt!]
\centering
\includegraphics[width=0.98\textwidth]{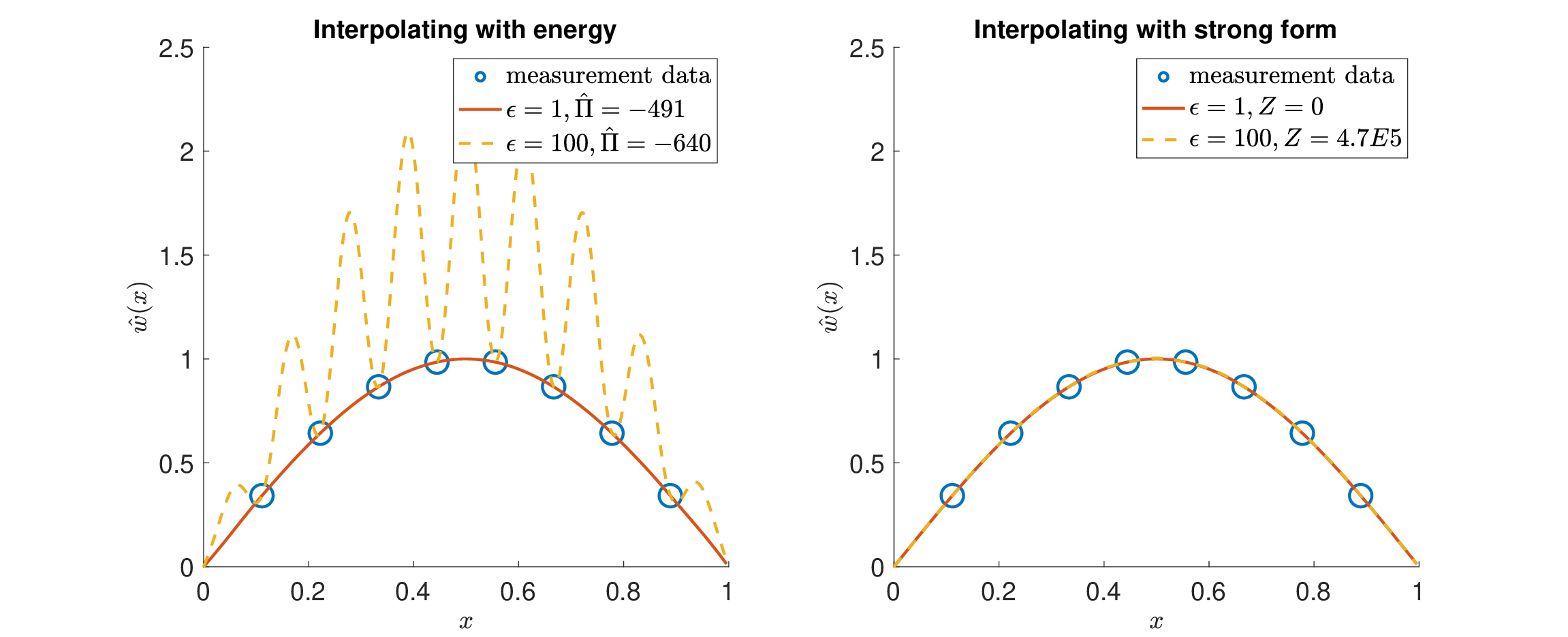}
\caption{ In this example, minimizing the energy $(\hat \Pi)$ over $\epsilon$ will lead to poor interpolations (left), whereas the minimum strong form loss (which we call $Z$, to distinguish it from the loss involving the strong form \textit{and} penalty in Eq. \eqref{L1}), is obtained when the physics parameter is correct, leading to good interpolations (right). It is interesting to note that energy and strong form loss produce very different interpolations on what appears to be the same problem.}
\label{pp3}
\end{figure}

  To illustrate this, consider the following problem. The measurement data is taken from $u(x) = \sin \pi x$, which corresponds to a source term of $s(x) = \pi^2 \sin \pi x$. We assume a parameterized source term of the form $b(x) = \epsilon \pi^2 \sin \pi x$. This means that the exact model should be recoverable with $\epsilon=1$. But, the optimization problem in Eq. \eqref{L5p} is unbounded and thus does not converge to a solution. Figure \ref{pp3} illustrates why this is the case. We fix the physics parameter $\epsilon$ and enforce the constraints with Lagrange multipliers. We can then compare the loss function values for the true physics parameters $\epsilon=1$, and a highly non-physical value of $\epsilon=100$. The same process is repeated on the strong form loss for the sake of comparison. The energy associated with the exact solution is greater than the energy associated with a highly non-physical interpolation of the data. This illustrates that finding the minimum of the variational energy over $\epsilon$ will not recover an approximation of the model that generated the data, since these non-physical solutions have lower energy. This simple example demonstrates an inconvenient fact---the formulation of the solution reconstruction problem depends quite delicately on the choice of physics loss. Despite seeming like a relatively minor modification of the notion of a solution to the BVP, the variational energy cannot be used as a loss function to reconstruct the solution in the same way that the strong form can. This shows that the penalty approach to solution reconstruction fails to satisfy the robustness criterion, as it is sensitive to the choice of physics loss.

\subsection{Data consistency}

  The physics parameters $\boldsymbol{\epsilon}$ are identifiable if they can be uniquely determined from the measurement data and a given parameterization of the physics \cite{maclaren_what_2020}. To motivate why data consistency is important for solution reconstruction, we will investigate whether solving Eq. \eqref{modelproblem} with the methods reviewed above necessarily leads to an identifiable problem. Let us assume that the body force is parameterized with the same set of basis functions as the displacement, i.e.,

\begin{equation} \label{body}
b(x;\boldsymbol{\epsilon}) = \sum_{j=1}^P \epsilon_j \sin( j \pi x).
\end{equation}
Note that $P \neq N$ in general. That the discretization enforces $b(0)=b(1)=0$ is immaterial in this problem. Parameterizing the missing physics with a flexible discretization of this sort is a common approach in the literature. For example, in \cite{he_physics-informed_2020}, a spatially varying material parameter is represented as a neural network and learned from the data. In \cite{zou_correcting_2024}---the work in which the assumption of consistency between the parameterized physics and the true model is relaxed---a neural network is used to represent a missing source term that the parameterization of the physics does not capture. Whereas prior knowledge informs the parameterized physical models shown in this work, the neural network discrepancy model in \cite{zou_correcting_2024} is general. Here, we consider that Eq. \eqref{body} is the equivalent of the discrepancy model, simply written as a sine series rather than a neural network. It is not important for the following arguments that there are no additional trainable parameters beyond those of the discrepancy model. We will use the strong form loss and enforce the equality constraints with Lagrange multipliers, which is simply Eq. \eqref{L2} applied to the model problem. The objective and optimality conditions are 

\begin{equation} \label{L2p}
\begin{aligned}
&\mathcal{L}_2 = \frac{1}{2}\int_0^1 \qty( \frac{\partial^2 \hat w}{\partial x^2} + b(x;\boldsymbol{\epsilon}) )^2 dx + \sum_{i=1}^C \lambda_i \Big( \hat w(x_i) - v_i \Big), \\
&\pd{\mathcal{L}_2}{\boldsymbol{\theta}} =\mathbf{0}, \quad  \pd{\mathcal{L}_2}{\boldsymbol{\epsilon}} = \mathbf{0}, \quad \pd{\mathcal{L}_2}{\boldsymbol{\lambda}} = \mathbf{0}.
\end{aligned}
\end{equation}

Plugging in the discretization of the displacement and the source term, factoring coefficients outside of the integral, taking derivatives, and organizing into a block matrix, the stationarity condition of Eq. \eqref{L2p} can be written as

\begin{equation} \label{sys}
    \mathbf{S} \begin{bmatrix} \boldsymbol{\theta} \\ \boldsymbol{\epsilon} \\ \boldsymbol{\lambda}
\end{bmatrix} = \begin{bmatrix} \mathbf{0} \\ \mathbf{0} \\ \mathbf{v}
\end{bmatrix},
\end{equation}

\noindent where $\mathbf{S}$ is the $[N+P+C] \times [N+P+C]$ matrix. Depending on the relationship between the number of terms in the displacement discretization $N$, the number of terms in the body force discretization $P$, and the number of constraints $C$, the matrix $\mathbf{S}$ may not be full-rank. In particular, if the number of terms in the body force discretization exceeds the number of constraints, the system will not have a unique solution. Non-unique solutions to Eq. \eqref{sys} correspond to non-identifiable physics parameters $\boldsymbol \epsilon$. Additionally, by forming the objective given in Eq. \eqref{L1} with the sine series discretization and writing out the optimality condition, a penalty method for the model problem of Eq. \eqref{modelproblem} also leads to non-identifiable physics parameters when the number of degrees of freedom in the parameterization exceeds the number of constraints. Owing to the large number of trainable parameters in the neural network discrepancy model, the methods introduced in \cite{zou_correcting_2024} are prone to this very shortcoming. Likely, the spectral bias of neural networks prevents the learning of high-frequency nullspace elements \cite{rahaman_spectral_2019}.

  When it is possible to form a linear system for the optimality condition, it is simple to assess the uniqueness of the solution through rank tests on the system matrix. However, when Eq. \eqref{L2p} is solved with iterative optimization techniques---as is typically the case in the machine learning literature---it may not be clear that one among a family of solutions is being found by the optimizer. This example shows that neglecting to match the number of parameters in the discrepancy model to the number of measurements can lead to non-identifiability of the reconstructed solution. Though we do not claim that respecting the data consistency condition guarantees identifiability and uniqueness, we do claim that failing to respect it will introduce non-identifiability and non-uniqueness to the solution reconstruction problem. These are characteristics of a problem we would like to avoid when possible.

  At this point, we have shown that standard PINNs-type methods do not provide an interpretable quality measure, that the energy loss is not a suitable objective for the formulations of the solution reconstruction problem discussed thus far, and that failing to respect the data consistency condition can lead to non-identifiability issues even for simple linear problems. We will now investigate why this is the case. This investigation will reveal further robustness issues with PINNs-type approaches---especially in the setting where the true and the parameterized model are inconsistent---and pave the way for a solution.


\section{Constraint forces}

  The discussion above has shown that if the energy objective is to be used, we need an auxiliary criterion to determine the physics parameter $\epsilon$. We will see that these considerations also lead us to a method that ensures data consistency and interpretability as well. Figure \ref{pp3} highlights that the constraints play less of a role in obtaining the solution with $\epsilon=1$ than with $\epsilon=100$. We can begin to formalize this intuition as follows: in the context of constrained optimization with Lagrange multipliers, a constraint is said to be ``inactive'' when it is naturally satisfied at the minimum of the objective. An inactive constraint has a corresponding Lagrange multiplier of $\lambda=0$. With the energy loss, when possible, we want to find the physics parameter $\epsilon$ such that the constraints are inactive. This would occur when the model that generated the data is exactly recovered, but this does not correspond to a minimum of the energy. An inactive constraint is analogous to the idea that a parameterization of the physics that is consistent with the measurement data will satisfy constraints in the context of a penalty method, even with a very small penalty parameter. What we have seen is that, when possible, minimizing the strong form loss finds parameters $\epsilon$ such that the constraints are inactive, but minimizing the energy loss does not. This is an interesting and crucial difference between the two methods. We can now demonstrate why, in the context of the model problem, the strong form loss has this property, but the energy loss does not.

  For the sake of exposition, we will work with the continuous version of the strong form loss of the solution reconstruction model problem, with the magnitude of the body force as the only parameter to be determined. This means that the solution is not discretized. There will be one constraint for simplicity. The Lagrange function corresponding to the strong form loss is

\begin{equation} \label{L2c}
    \mathcal{L}_2\Big(\hat w(x),\epsilon,\lambda\Big) = \frac{1}{2}\int_0^1 \qty( \frac{\partial^2 \hat w}{\partial x^2} + \epsilon b(x) )^2 dx + \lambda\Big( \hat w(x_c) - v \Big),
\end{equation}

\noindent where $x_c$ is the position of the single constraint. The governing equation for the physics is obtained by taking the variation with respect to the displacement. This reads

\begin{equation*} \label{variation1}
    \delta_w \mathcal{L}_2 = \int_0^1 \qty( \frac{\partial^2 \hat w }{\partial x^2} + \epsilon b(x) )\frac{\partial^2 \delta \hat w}{\partial x^2} dx + \lambda \delta \hat w(x_c) = 0.
\end{equation*}

We will attempt an interpretation of the Lagrange multiplier by bringing it inside the integral as a source term. This will prove useful later in understanding the properties of the strong form loss. This can be accomplished as follows

\begin{equation*}
    \lambda \delta \hat w(x_c) = \lambda \int_0^1  \delta \hat w(x) \delta(x-x_c) dx = \lambda \int_0^1 \delta \hat w(x) \frac{\partial ^2}{\partial x^2} I(x-x_c) dx,
\end{equation*}

\noindent where $\delta(x-x_c)$ is a Dirac-delta function centered at $x_c$ and $I(x-x_c)$ is defined as 

\begin{equation*}
I (x-x_c) = 
    \begin{cases}
        0, & x < x_c, \\
        x-x_c, &  x \geq x_c.
    \end{cases}
\end{equation*}

This allows us to integrate by parts the derivatives onto the variation $\delta \hat w$. Our model problem has homogeneous Dirichlet boundaries, so $\delta \hat w(0) = \delta \hat w(1) = 0$. Integrating by parts twice, we obtain

\begin{equation*}
     \lambda \delta \hat w(x_c) = \lambda \Big( \int_0^1 I(x-x_c) \frac{\partial^2 \delta \hat w}{\partial x^2} dx - I(x-x_c) \pd{\delta \hat w}{x} \Bigg |_0^1 \Big).
\end{equation*}

We have asserted that the boundary conditions are known deterministically, meaning that any constraint on the solution is interior to the domain. This means that $x_c>0$, allowing us to simplify the second term to

\begin{equation*}
   - \lambda I(x-x_c) \pd{\delta \hat w}{x} \Bigg|^1_0 = -\lambda (1-x_c) \pd{\delta \hat w}{x}(1).
\end{equation*}

Finally, it can be shown with integration by parts that

\begin{equation*}
    \int_0^1 x \frac{\partial^2 \delta \hat w}{\partial x^2} dx = \pd{\delta \hat w}{x}(1).
\end{equation*}

This allows us to write the variation of the Lagrange function as 

\begin{equation} \label{final}
    \delta_w \mathcal{L}_2 = \int_0^1 \qty( \frac{\partial^2 \hat w}{\partial x^2} + \epsilon b(x) + \lambda \Big(  I(x-x_c) - (1-x_c) x \Big) ) \frac{\partial^2 \delta \hat w}{\partial x^2}dx = 0.
\end{equation}

In the case of our elastic model problem with zero boundaries, the Lagrange multiplier introduces a ``constraint force'' in the form of a hat function centered at $x_c$, which is zero at the two boundaries. Eq. \eqref{final} states that, at the solution, we have 

\begin{equation}\label{condense}
    \frac{\partial^2 \hat w}{\partial x^2} + \epsilon b(x) + \lambda \Big(  I(x-x_c) - (1-x_c) \Big) = 0.
\end{equation}

Eq. \eqref{condense} can be used to treat the displacement as a function of the physics parameter and the Lagrange multiplier. This allows us to reformulate the stationarity condition of Eq. \eqref{L2c} as

\begin{equation}\label{mcfp}
\begin{aligned}
   \underset{\epsilon}{\text{argmin }} \frac{\lambda^2}{2} \int_0^1 \Big(   I(x-x_c) - (1-x_c)  \Big)^2 dx,\\
   \text{s.t. } \hat w(x_c) - v =0, \quad \frac{\partial^2 \hat w}{\partial x^2} + \epsilon b(x) + \lambda \Big(  I(x-x_c) - (1-x_c) \Big) = 0.
\end{aligned}
\end{equation}


Eq. \eqref{mcfp} shows that minimizing the strong form residual can be recast as minimization of the Lagrange multiplier. The Lagrange multiplier is treated as a function of the physics parameter through the constraints. This demonstrates why the strong form loss finds $\epsilon$ such that the constraint is inactive in the special case that the data can be exactly reproduced by the parameterized source term. When the data is inconsistent with the parameterization of the physics, the Lagrange multipliers will apply forces to the system to enforce the constraints. Though the constraints on the displacement can be satisfied by choosing the Lagrange multiplier appropriately for any physics parameter $\epsilon$, Eq. \eqref{mcfp} shows that we find $\epsilon$ such that the constraint force is minimal. Thus, analysis of the model problem has illustrated the following important points:

\begin{itemize}
    \item The optimal choice of physics parameters $\epsilon$ is such that the constraint force, and hence Lagrange multiplier, is minimal;
    \item By controlling the magnitude of the constraint force, the Lagrange multiplier itself acts as an interpretable quality measure which gauges how consistent the parameterization of the physics is with the measured system;
    \item Constraints add source terms to the system when the measurement data cannot be reproduced by an appropriate choice of $\epsilon$;
    \item Determining the nature of these source terms is problem-dependent and requires non-trivial calculations to bring the Lagrange multiplier term inside the system residual.
\end{itemize}

  The model problem drives home a rather general, yet intuitive idea: when the constraint term from Lagrange multipliers can be interpreted as a force, the physics parameters $\epsilon$ should be such that the force of constraint is minimal. We have shown that in the context of our model problem, minimizing the strong form loss enforces this condition automatically. The reason that the energy loss does not work to interpolate the data is that minimizing the energy over $\epsilon$ does not correspond to minimizing the constraint forces. The interpretation of the Lagrange multiplier as a fictitious source term given in Eq. \eqref{final} depends on the interaction between the Lagrange multiplier term and the physics loss. We can interpret constraint force with the energy loss by writing out the Lagrange function corresponding to the variational energy and a single constraint:

\begin{equation*} \label{abc}
\begin{aligned}
    \mathcal{L}_5 = \int_0^1 \frac{1}{2} \qty(\pd{\hat w}{x})^2 - \epsilon b(x) \hat w(x) dx + \lambda\Big( \hat w(x_c) - v \Big) \\= \int_0^1 \frac{1}{2} \qty(\pd{\hat w}{x})^2 - \epsilon b(x) \hat w(x) + \lambda \hat w(x) \delta(x-x_c) dx  - \lambda v.
\end{aligned}
\end{equation*}

The Lagrange multipliers act as point forces in the context of the energy loss. The stationarity condition for the energy provides the governing equation for the physics:

\begin{equation} \label{stat}
    \delta \mathcal{L}_5 = \int_0^1 \pd{\hat w}{x}\pd{\delta \hat w}{x} - \epsilon b(x) \delta \hat w + \lambda \delta(x-x_c) \delta \hat w dx = 0.
\end{equation}

We cannot integrate by parts the spatial derivative from the test function because the second derivative of the displacement is not defined under the action of a point force. In practice, the space of variations $\delta \hat w$
is discretized and Eq. \eqref{stat} is satisfied for each basis function. The minimum constraint force principle, which is implicit in the strong form loss, is not built into the energy loss. Working with the discretized displacement, the minimum constraint force principle with our single constraint elastic model problem can be written as

\begin{equation} \label{mcf}
\begin{aligned}
   & \underset{\epsilon}{\text{argmin }} \frac{1}{2} \lambda(\epsilon)^2 \\
   & \text{s.t. } \pd{\mathcal{L}_5}{\boldsymbol{\theta}} = \mathbf{0}, \quad \pd{\mathcal{L}_5}{\lambda} = 0.
\end{aligned}
\end{equation}

\begin{figure}[hbt!]
\centering
\includegraphics[width=0.9\textwidth]{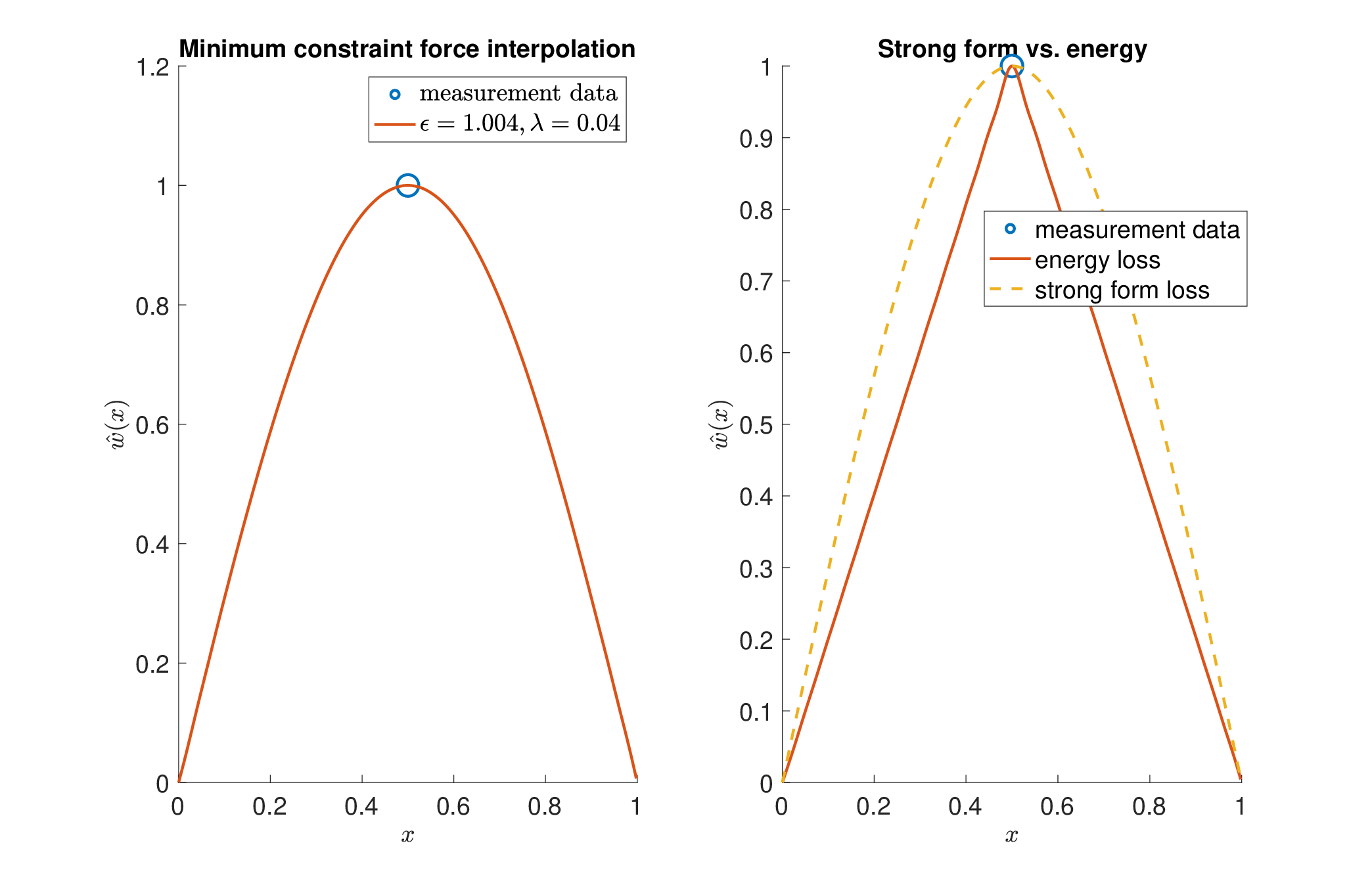}
\caption{Remedying the issues with the energy objective by using the minimum constraint force method (left). The minimum constraint force interpolation now accurately recovers the physics parameter and satisfies the constraint with a small constraint force ($\lambda$). The strong form and energy loss lead to different interpolations when driven only by constraint forces (right).}
\label{pp4}
\end{figure}

Eq. \eqref{mcf} finds the stationary point of the energy, which corresponds to the minimum norm of the constraint force. This is discretized version of Eq. \eqref{mcfp}, but with the energy objective as opposed to strong form loss. For a given $\epsilon$, the Lagrange objective function is formed in the ``inner-loop'' problem, which determines the Lagrange multiplier $\lambda$ and displacement parameters $\boldsymbol{\theta}$. The outer-loop problem updates the physics parameter such that the constraint force, captured by the Lagrange multiplier, is minimized. To the best of the authors' knowledge, the minimum constraint force principle is a novel way of formulating the solution reconstruction problem in the machine learning literature. Revisiting the problem shown in Figure \ref{pp3}, we can show that the minimum constraint force principle discovers the true physics parameter, thus giving rise to the desired interpolation of the data. This is shown for the model problem in Figure \ref{pp4}.

  For the 1D second-order linear model problem with a single unknown physics parameter and a single constraint, we have shown that Lagrange multipliers behave like constraint forces with a spatial form dictated by the choice of physics loss. It is also possible to back out an interpretation of the constraint forces coming from the weak form objective and Lagrange multipliers with this same problem. This is shown in Appendix A. We have shown that the constraint force is ``global'' for the strong form loss, meaning that it is a source term spread over the whole domain. On the other hand, the constraint force corresponding to the energy loss was a localized point force. Thus, owing to the different constraint forces, different formulations of the physics loss lead to different reconstructed solutions. In the case that the parameterization of the source term is inaccurate, the solution will be primarily driven by constraint forces. Figure \ref{pp4} shows the energy and strong form loss with no body force used to interpolate a single measurement at the center of the domain. As expected, the distributed constraint force from the strong form loss yields smoother interpolations than the concentrated constraint force of the energy loss. Thus, inconsequential choices about how to formulate the BVP interact with the Lagrange multiplier method for constraint enforcement in interesting and unpredictable ways. The different smoothness properties of the reconstructed solutions coming from different physics losses are another example of how the methods we have discussed thus far are not robust. A similar analysis can be carried out to show that penalty methods introduce constraint forces which are dictated by the formulation of the physics loss. As shown in Appendix B, the penalty methods have the character of a displacement-dependent source term, which means that they influence the solution operator as well. In general, this is an undesirable property, as the physics of the problem is tampered with in more fundamental ways. Thus, in addition to not being interpretable, the penalty formulation of the strong form loss objective suffers from the same robustness issues as a result of the introduction of constraint forces and their influence on the solution operator.


  We have shown the importance of data consistency as a preventative measure against non-identifiability. We showed that penalty and ``out of the box'' Lagrange multiplier methods are not interpretable, in the sense that they do not provide a quality measure of the reconstructed solution with physical units. Then, we showed that the formulation of the physics loss has a significant impact on the solution reconstruction problem. In particular, the energy loss could not be inserted into the same sort of optimization problem as the strong form loss. This motivated a detour into Lagrange multipliers in the context of the model problem, where we showed that a Lagrange multiplier can be interpreted as the magnitude of a constraint force whose spatial form is dictated by the physics loss. Though this analysis was carried out for a single constraint, it extends to the case of multiple constraints. The variegated constraint forces surfaced another robustness issue, namely that constraint-driven reconstructions will exhibit differences in accordance with the choice of physics loss. \textit{Reconstructions which are driven by constraints are exactly the situation we expect when the parameterized model is inconsistent with the true physical system, as satisfying the data constraints is not possible for any choice of physics parameters.} The minimum constraint force principle was proposed as a criterion that underlies both the strong form and energy losses. Though the strong form loss satisfies this principle automatically, the energy loss needed to be supplemented by the minimum constraint force principle in order to be usable.

  We now propose a method that remedies the issues we have identified. We wish to formulate a solution reconstruction method that is data-consistent by construction, which provides an interpretable quality measure, and which produces similar reconstructed solutions for any choice of physics loss, even when the reconstruction is driven by constraint forces. This will be called the ``explicit constraint force method'' (ECFM).


\section{Explicit constraint force method (ECFM)}

  It was possible to obtain the exact form of the constraint forces for the strong form loss and variational energy with the 1D linearly elastic model problem. For more complex physics, boundary conditions, and geometries---though similar concepts apply---this becomes intractable. Thus, in order to avoid the unpredictable effects of interactions between the physics loss and the technique for constraint enforcement, it is desirable to explicitly control the constraint forces. Doing so will lead to reconstructions which are robust with respect to the choice of physics loss and technique for constraint enforcement. With slight abuse of terminology, we will refer to source terms as ``forces'' even in the context of heat transfer. For the model problem, the source term introduced from Lagrange multipliers was in fact a distributed force, but in heat transfer, it would be a volumetric heat source. The source terms used to enforce constraints in elasticity and heat transfer will both be referred to as ``constraint forces.'' 

  As previously stated, we also seek a solution reconstruction method which is automatically data-consistent and provides a quality measure of the reconstruction. The three desiderata can be obtained by looking beyond Lagrange multipliers and penalty methods for constraint enforcement. Our method consists of two main steps---the first step is to introduce additional source terms which are centered at the locations of the constraints. In the second step, we treat the solution parameters and constraint force magnitudes as dependent on the physics parameters. Working first with equality constraints, the strong form of the abstract BVP is 

\begin{equation} \label{gamma}
\begin{aligned}
    &\mathcal{G}\Big( \mathbf{w}(\mathbf{x}) ; \boldsymbol{\epsilon}_1 \Big) + \mathbf{b}(\mathbf{x};\boldsymbol{\epsilon}_2) + \sum_{i=1}^C \boldsymbol{\lambda}_i \Gamma( \mathbf{x} - \mathbf{x}_i) = \mathbf{0}  \\
    &\text{s.t. } \mathbf{w}(\mathbf{x}_i) - \mathbf{v}_i = \mathbf{0} \quad i=1,\dots,C.
\end{aligned}
\end{equation}

Note that the constraints cannot be enforced with Lagrange multipliers or the penalty method without introducing fictitious source terms to the system whose form is outside of the analyst's control. The goal is to enforce the constraints exclusively through these new source terms. We will give examples of how to enforce the constraints without penalties or Lagrange multipliers shortly. The scalar function $\Gamma(\mathbf{x}-\mathbf{x}_i)$ can be chosen by the analyst based on desired smoothness properties of the reconstruction or prior knowledge of the missing physics. If there is no prior knowledge guiding the choice of the constraint forces, the most parsimonious choice would be a polynomial basis with local support. The parameters $\boldsymbol{\lambda}_i$ are no longer Lagrange multipliers, but rather the unknown magnitudes of the constraint forces. By explicitly introducing new source terms into the system and avoiding Lagrange multipliers or penalty methods, we gain control of the constraint forces. We want to avoid a situation as in Figure \ref{pp4}, where different physics losses have different constraint forces and thus yield different reconstructed solutions.

  Eq. \eqref{gamma} can be satisfied for any choice of $\boldsymbol{\epsilon} = [\boldsymbol{\epsilon}_1,\boldsymbol{\epsilon}_2]^T$. The optimal choice of physics parameters should satisfy the minimum constraint force principle. Given that the constraint forces need not be point forces, the total constraint force is 

\begin{equation} \label{mag}
\begin{aligned}
    z(\boldsymbol{\epsilon}) = \frac{1}{2} \int \Big \lVert \sum_{i=1}^C \boldsymbol{\lambda}_i(\boldsymbol{\epsilon}) \Gamma(\mathbf{x}-\mathbf{x}_i) \Big \rVert^2 d\Omega =\frac{1}{2} \sum_{i=1}^C \sum_{j=1}^C \boldsymbol \lambda_i \cdot \boldsymbol \lambda_j \qty(\int\Gamma(\mathbf{x}-\mathbf{x}_i) \Gamma(\mathbf{x}-\mathbf{x}_j) d\Omega) \\:=\frac{1}{2} \sum_{i=1}^C \sum_{j=1}^C \boldsymbol \lambda_i \cdot \boldsymbol \lambda_j H_{ij}
\end{aligned}
\end{equation}

Note that we can normalize the magnitude of the constraint force functions such that the diagonal elements of $\mathbf{H}$ are unity. The reconstructed solution is then obtained by minimizing the total constraint force in terms of the physics parameters, where the displacement and the coefficients on the constraint force functions are determined for given physics parameters through Eq. \eqref{gamma}. This is the inner-loop problem, which corresponds with the following outer-loop optimization problem:

\begin{equation}
    \underset{\boldsymbol{\epsilon}}{\text{argmin }} \frac{1}{2} \boldsymbol{\lambda} : \mathbf{H} \boldsymbol{\lambda}.
\end{equation}

\noindent where $\boldsymbol \lambda = \boldsymbol \lambda_{C \times D}$ puts the constraint force vectors corresponding to each constraint position in rows ($D$ is the number of spatial dimensions). One way to solve Eq. \eqref{gamma} without Lagrange multipliers or a penalty method is to construct the weak form. The weak form of the parameterized BVP with a displacement discretized by parameters $\boldsymbol \theta$ reads

\begin{equation*} \label{weakres}
    \hat R_j = \int \mathcal{G}\Big( \mathbf{\hat w}(\mathbf{x};\boldsymbol{\theta}) ; \boldsymbol{\epsilon}_1 \Big) \cdot \mathbf{f}_j + \mathbf{b}(\mathbf{x};\boldsymbol{\epsilon}_2) \cdot \mathbf{f}_j + \sum_i \boldsymbol{\lambda}_i \Gamma(\mathbf{x}-\mathbf{x}_i) \cdot \mathbf{f}_j d\Omega = 0, \quad j=1,\dots,M,
\end{equation*}

\noindent for a given set of test functions $\{\mathbf{f}_j(\mathbf{x})\}_{j=1}^M$. Given a fixed $\boldsymbol{\epsilon}$, this is a system of equations involving the unknown solution parameters $\boldsymbol{\theta}$ and constraint force magnitudes $\boldsymbol{\lambda}$. The constraint is appended to the system of equations, which means that the governing equation for the inner-loop problem is 

\begin{equation} \label{res}
    \mathbf{R}(\boldsymbol{\theta},\boldsymbol{\lambda} | \boldsymbol{\epsilon}) = 
    \begin{bmatrix}
        \mathbf{\hat R}(\boldsymbol{\theta},\boldsymbol{\lambda}|\boldsymbol{\epsilon})
        \\ \mathbf{\hat w}(\mathbf{x};\boldsymbol{\theta}) - \mathbf v
    \end{bmatrix} = \mathbf{0}.
\end{equation}

We use the notation $\mathbf{R}(\boldsymbol \theta, \boldsymbol \lambda|\boldsymbol \epsilon)$ to state explicitly that the constrained system is solved for given physics parameters $\boldsymbol \epsilon$. A similar system of equations can be constructed by writing out the optimality condition for the minimum residual of the strong form loss of Eq. \eqref{gamma}. Because the constraint forces are introduced explicitly, the choice of physics loss will have no bearing on them. ECFM for solution reconstruction can be written compactly as

\begin{equation} \label{problem}
    \begin{aligned}
      &  \underset{\boldsymbol{\epsilon}}{\text{argmin }} \frac{1}{2} \boldsymbol{\lambda} : \mathbf{H} \boldsymbol{\lambda} \\
      &  \text{s.t. } \mathbf{R}(\boldsymbol{\theta},\boldsymbol{\lambda}|\boldsymbol{\epsilon}) = \mathbf{0}.
    \end{aligned}
\end{equation}

\noindent where ``$:$'' indicates a contraction on both indices of a matrix. There are many methods to solve constrained optimization problems of this sort, and ECFM does not depend on a particular approach. In examples of Section 7, we will specify the optimization method on a case-by-case basis. Once the physics parameters are obtained from this optimization problem, the parameters $\boldsymbol \theta$ which control the reconstructed solution are obtained from the solution of the inner-loop problem. The quality of the reconstructed solution is assessed by the magnitude of the total constraint force $z$ using Eq. \eqref{mag}. The total constraint force provides insight into the extent of additional ``forcing'' required to make the parameterized system consistent with the measurement data. Because it can be interpreted in the same units as any other source term---forces in elasticity, heat fluxes for a thermal problem---it provides an interpretable means of assessing the quality of the reconstructed solution. Finally, because the constraint force functions are centered on the constraint positions, there are as many additional degrees of freedom introduced to the system as there are constraints. This avoids data-inconsistency issues encountered with a more general discrepancy model by ensuring that Eq. \eqref{res} always has a unique solution. We will now show the efficacy of this approach with three numerical experiments from systems in elasticity and heat transfer.


\section{Numerical examples}

\subsection{1D elastic bar}

  We revisit the model problem given in Eq. \eqref{modelproblem} to show that the explicit constraint force method provides reconstructed solutions which are interpretable, robust, and identifiable (for this particular problem) by virtue of being data consistent. To this end, we will show that the total constraint force given by Eq. \eqref{mag} provides a convenient quality measure for the reconstructed solution, that all three physics losses provide equivalent interpolations of the measurement data, and that we obtain a unique reconstructed solution. We will assume that we have equality constraints on the data (zero measurement noise) and that the solution is discretized with the sine series of Eq. \eqref{fourier}. Using this discretization, the equality constraints can be written as  

\begin{equation} \label{eqcon}
\begin{aligned}
    \sum_{j=1}^N \theta_j f_j(x_i) = v_i, \quad i=1,2,\dots,C \quad 
    \text{or equivalently}\quad \mathbf{G} \boldsymbol{\theta} - \mathbf{v} = \mathbf{0}, \\ \quad \text{where }\quad G_{ij} = f_j(x_i).
\end{aligned}
\end{equation}

The shape functions are the sine series $f_j(x) = \sin( j \pi x)$. By writing out the strong form loss with the constraint force functions introduced per Eq. \eqref{gamma}, the governing system of equations is obtained by taking its gradient with respect to the solution parameters. This reads

\begin{equation*}
    \hat R_k^{\text{sf}} = \int_0^1 \qty( \sum_{j=1}^N \theta_j \frac{\partial^2 f_j}{\partial x^2} + \epsilon b(x) +\sum_{i=1}^C \lambda_i \Gamma(x-x_i) ) \frac{\partial^2 f_k}{\partial x^2 } dx = 0, \quad k=1,2,\dots,N.
\end{equation*}

The superscript ``sf'' indicates that this system of equations comes from the strong form loss. The constraint force functions $\Gamma(x-x_i)$ are taken as piecewise linear hat functions centered at the constraint location $x_i$, which respect the interpolation property on the measurement grid. We can write this in matrix-vector form more concisely as

\begin{equation*}
    \mathbf{\hat R}^{\text{sf}}(\boldsymbol{\theta},\boldsymbol{\lambda}|\epsilon) = \mathbf{\tilde K} \boldsymbol{\theta} + \epsilon \mathbf{\tilde F} + \boldsymbol{\tilde \Gamma} \boldsymbol{\lambda} = \mathbf{0}.
\end{equation*}

The equality constraints on the displacement are appended to this system, and the full system of equations is written as 

\begin{equation}\label{sys1}
    \mathbf{R}^{\text{sf}}(\boldsymbol{\theta},\boldsymbol{\lambda}|\epsilon) = \begin{bmatrix} \mathbf{\tilde K} & \boldsymbol{\tilde \Gamma} \\ \mathbf{G} & \mathbf{0}
    \end{bmatrix} \begin{bmatrix} \boldsymbol{\theta} \\ \boldsymbol{\lambda} \end{bmatrix} - \begin{bmatrix} -\epsilon \mathbf{\tilde F} \\ \mathbf{v}
    \end{bmatrix} = \mathbf{0}.
\end{equation}

Note that this system is square by construction and therefore avoids issues with non-uniqueness in the reconstructed solution. Per Eq. \eqref{problem}, the solution using the explicit constraint force method with the strong form loss is given by 

\begin{equation} \label{sf1}
\begin{aligned}
   & \underset{\epsilon}{\text{argmin }} \frac{1}{2} \boldsymbol{\lambda} \cdot \mathbf{H} \boldsymbol{\lambda} \\
   & \text{s.t. } \mathbf{R}^{\text{sf}}(\boldsymbol{\theta},\boldsymbol{\lambda}|\epsilon) =  \mathbf{0}.
\end{aligned}
\end{equation}

The weak form of the governing equation is taken by integrating against a set of suitable test functions per Eq. \eqref{weak}. The Bubnov-Galerkin weak form of the model problem is

\begin{equation*}
    \hat R_k^{\text{wf}} = \int_0^1 \qty( \sum_{j=1}^N \theta_j \frac{\partial^2 f_j}{\partial x^2} + \epsilon b(x) +\sum_{i=1}^C \lambda_i \Gamma(x-x_i) ) f_k dx = 0, \quad k=1,2,\dots,N.
\end{equation*}

The superscript ``wf'' is included to indicate it is the governing system from the weak form of the governing equation. This gives rise to an analogous system of equations given by 

\begin{equation}\label{sys2}
    \mathbf{R}^{\text{wf}}(\boldsymbol{\theta},\boldsymbol{\lambda}|\epsilon) = \begin{bmatrix} \mathbf{K} & \boldsymbol{\Gamma} \\ \mathbf{G} & \mathbf{0}
    \end{bmatrix} \begin{bmatrix} \boldsymbol{\theta} \\ \boldsymbol{\lambda} \end{bmatrix} - \begin{bmatrix} \epsilon \mathbf{F} \\ \mathbf{v}
    \end{bmatrix} = \mathbf{0}.
\end{equation}

Per Eq. \eqref{problem}, the solution using the explicit constraint force method with the weak form loss is then given by 

\begin{equation} \label{wff1}
\begin{aligned}
   & \underset{\epsilon}{\text{argmin }} \frac{1}{2} \boldsymbol{\lambda} \cdot \mathbf{H} \boldsymbol{\lambda} \\
   & \text{s.t. } \mathbf{R}^{\text{wf}}(\boldsymbol{\theta},\boldsymbol{\lambda}|\epsilon) =  \mathbf{0}.
\end{aligned}
\end{equation}

Note that the Bubnov-Galerkin weak form and the stationarity condition of the energy coincide. Thus, we have that $\mathbf{R}^{\text{wf}}(\boldsymbol{\theta},\boldsymbol{\lambda}|\epsilon) = \mathbf{R}^{\text{e}}(\boldsymbol{\theta},\boldsymbol{\lambda}|\epsilon)$, where the superscript ``e'' indicates the system of equations coming from the condition of stationarity of the variational energy. Because $\epsilon$ is obtained with the minimum constraint force principle, it is no longer necessary to distinguish between the weak form and the energy formulation. This was not the case, for example, when we sought $\epsilon$ by minimizing the energy. The constrained optimization problems of Eqs. \eqref{sf1} and \eqref{wff1} are solved by inverting the systems given by Eqs. \eqref{sys1} and \eqref{sys2} to obtain $\boldsymbol \lambda$ as an explicit function of the physics parameters $\boldsymbol \epsilon$. The gradients are computed per Appendix C and supplied to a standard quasi-Newton optimizer to find a minimum. 

\begin{figure}[hbt!]
\centering
\includegraphics[width=0.98\textwidth]{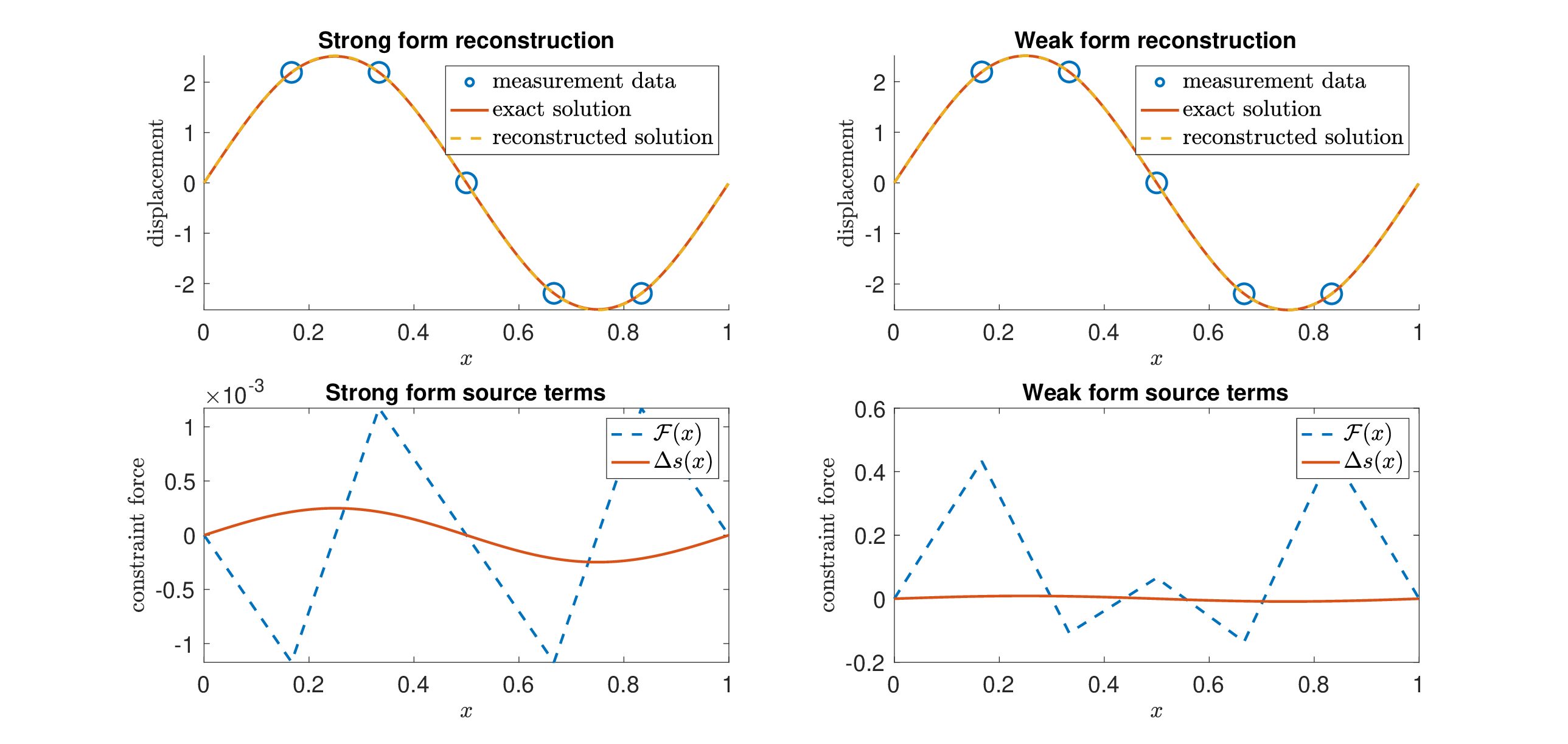}
\caption{Comparing the performance of the strong and weak formulations with the explicit constraint force method in the case that the true model can be exactly recovered. By not working with the strong form residual directly, the weak form incurs larger errors in the recovered constraint force than the strong form.}
\label{pp5}
\end{figure}

  We first demonstrate that Eqs. \eqref{sf1} and \eqref{wff1} accurately recover the model that generated the data when possible. We assume a manufactured solution of $u(x) = 100 \sin(2\pi x)/ 4\pi^2$, which corresponds to a source term of $s(x)=100 \sin(2\pi x)$. We take $C=5$ noiseless measurements of the solution on a uniform grid to generate the constraints $v_i$. The body force is parameterized as $b(x;\epsilon) = \epsilon \sin (2\pi x)$. We define $\mathcal{F}(x) = \sum_i \lambda_i \Gamma(x-x_i)$ as the distributed constraint force, and $\Delta s(x) = s(x) - \epsilon b(x)$ as the discrepancy between the true source term and the recovered source term. See Figure \ref{pp5} for the results. The explicit constraint force method finds a physics parameter of $\epsilon=99.99$ for both the strong and weak form methods. The constraint force objective given by Eq. \eqref{mag} is $z=2\text{e-}7$ in the case of the strong form and $z=0.020$ for the weak form. Given that these force values are orders of magnitude smaller than the applied forces to the system, this shows that the constraints are effectively inactive, and the solution is reconstructed entirely from fitting the source term.

\begin{figure}[hbt!]
\centering
\includegraphics[width=0.98\textwidth]{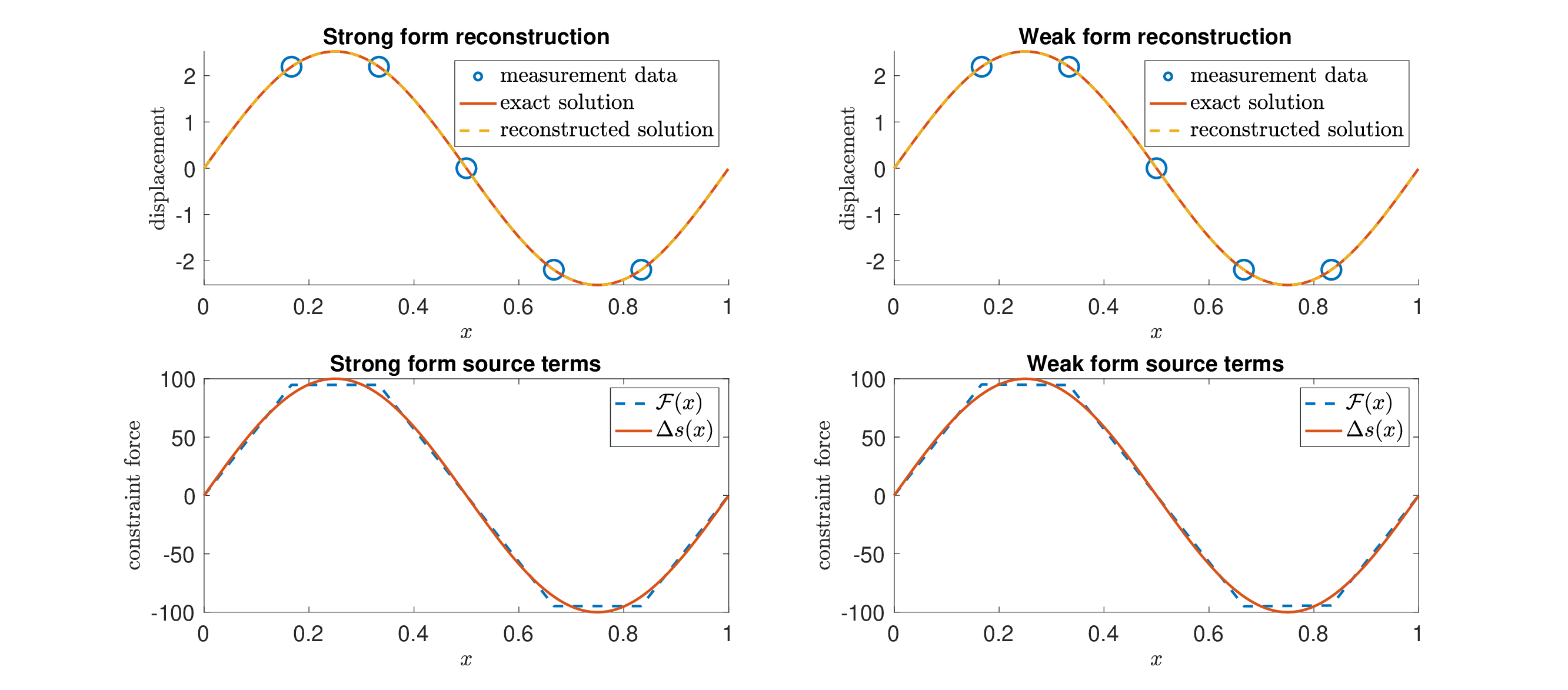}
\caption{Comparing the performance of the strong and weak formulations in a constraint-driven interpolation of the measurement data. Large constraint forces indicate that the parameterization of the missing physics is inaccurate. The constraint force distribution approximates the discrepancy between the true and parameterized source term, giving the analyst hints as to how to improve the model.}
\label{pp6}
\end{figure}

\begin{figure}[hbt!]
\centering
\includegraphics[width=0.45\textwidth]{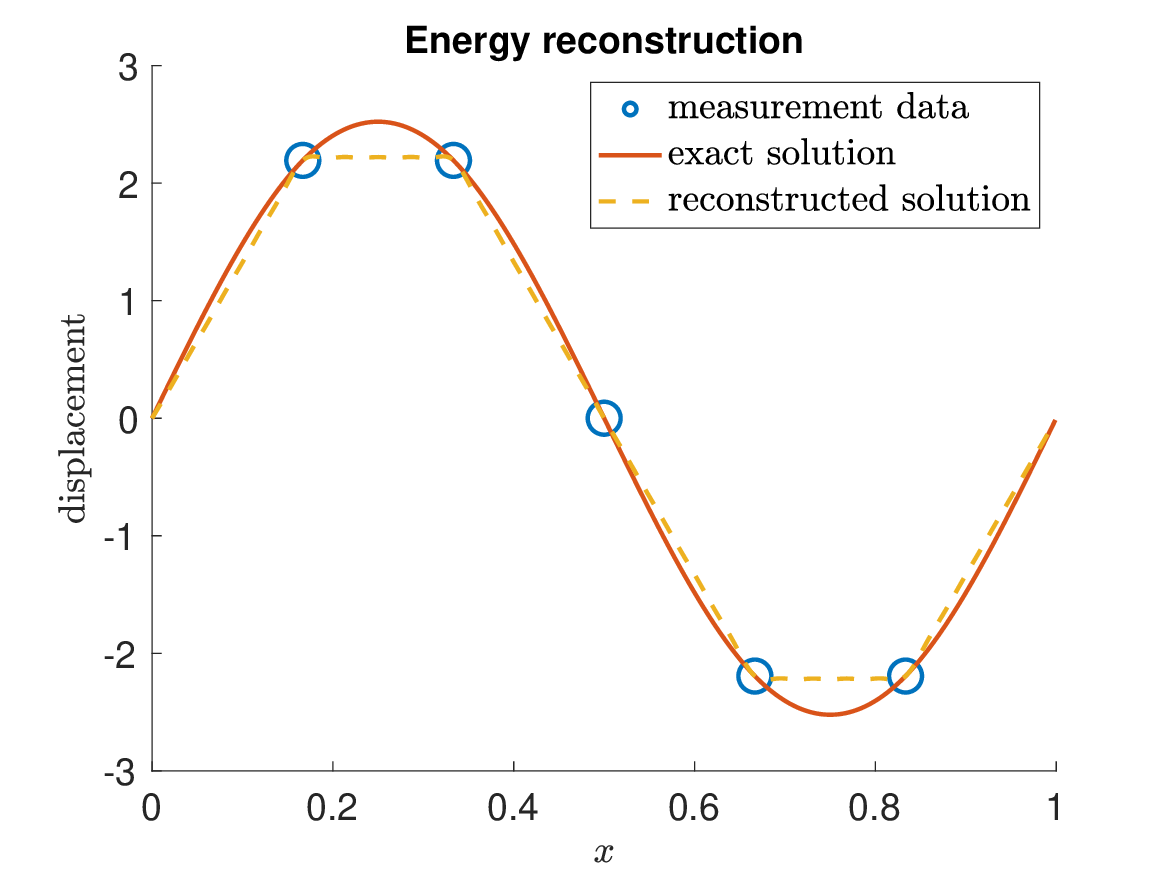}
\caption{Constraint forces using the energy loss are point forces, which leads to a solution with kinks. This problem was the original motivation for controlling the constraint forces.}
\label{extra}
\end{figure}

  Next, we can show that ECFM leads to reconstructed solutions with an interpretable quality metric that is robust to the choice of physics loss, even when the parameterized model is inconsistent with the measurement data. We assume the same manufactured solution of $u(x)=100 \sin( 2 \pi x )/4\pi^2$, which corresponds to a source term of $s(x) = 100 \sin(2 \pi x)$. We take $C=5$ noiseless measurements of the solution on a uniform grid to obtain the constraints $v_i$. The body force is parameterized as $b(x;\epsilon) = \epsilon \sin(\pi x)$. Note that it is no longer possible to recover the exact model, and we expect there to be a non-zero total constraint force $z$ applied to the system to enforce the equality constraints. In fact, because the parameterized body force is orthogonal to the true source term, we expect the optimizer to zero the physics parameter $\epsilon$. See Figure \ref{pp6} for the results. The physics parameter is $\epsilon=0.097$ for the strong form loss, and $\epsilon=9\text{e-}6$ for the weak form. Given the magnitude of the source term required to obtain the assumed displacements, these are extremely small values. This means that the reconstructed solution is driven entirely by constraint forces, which led to problems previously. Because the constraint forces are explicitly controlled, the two solution reconstruction methods are in good agreement. The constraint force magnitude is $z=2.5\text{e}3$ in both cases, providing a clear indication that the reconstructed solution is driven by the constraint forces. This suggests to the analyst that the parameterization of the missing physics is inaccurate. Contrast these reconstructed solutions with Figure \ref{extra}, where we use the energy objective with the minimum constraint force principle to obtain a reconstruction. Because the constraint force is not explicitly controlled, the reconstruction is unexpectedly non-smooth as a result of being driven by point forces. Though this may be desired in some cases, the advantage of ECFM is that the nature of the constraint force is of the analyst's choosing, rather than an unintended consequence of the physics loss. This demonstrates the robustness issue with PINNs-based approaches, where failing to enforce constraints with explicit source terms can lead to problematic reconstructions of the solution.

  Finally, we see that the constraint force provides an approximation of the missing physics. The difference between the real and parameterized source is approximated by the constraint force. This is another benefit of the explicit constraint force method---the constraint force can provide a hint as to the nature of the missing physics so that the parameterized model can be refined. This is similar to \cite{zou_correcting_2024}, except we do not have issues with data-inconsistency, which arises when the discrepancy model has more parameters than there are measurements. In our case, the richness of the discretization of the missing physics is matched to the informativeness of the measurement data. We see piecewise linear hat functions or similar as the most parsimonious choice of constraint force, introducing the fewest assumptions about the nature of the missing physics. There can be no oscillations of the constraint force between measurement data points, for example. We note that the ECFM is also convenient in explicitly using the discrepancy model to enforce the constraints.

\begin{figure}[hbt!]
\centering
\includegraphics[width=0.98\textwidth]{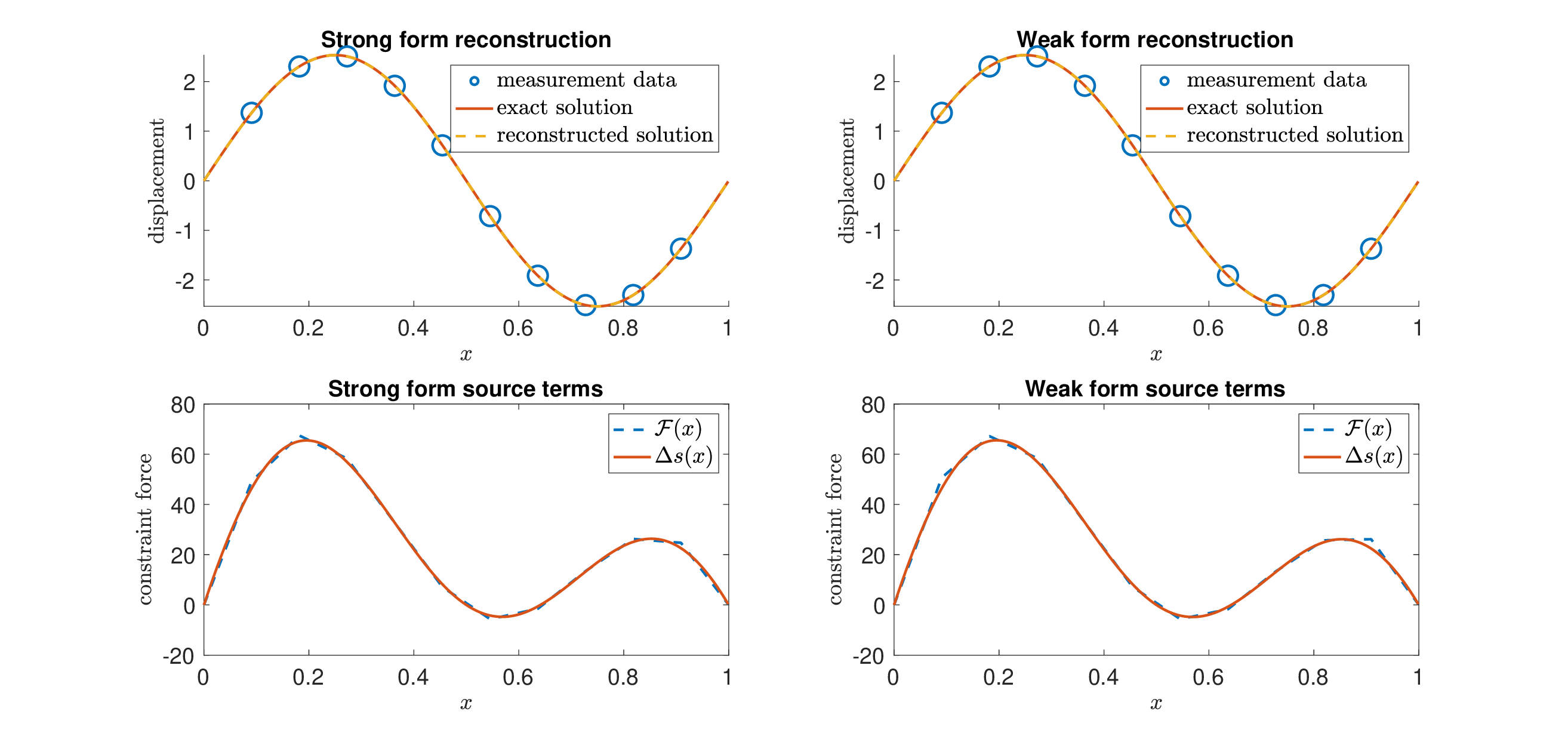}
\caption{Comparing the performance of the strong and weak formulations of the reconstructed solution with a partially accurate parameterization of the missing physics. This example demonstrates how improving the parameterization decreases the constraint force which enforces the constraints. We also see that increasing the number of constraints improves the accuracy of the constraint force's interpolation of the discrepancy between the real and parameterized source terms.}
\label{pp7}
\end{figure}

  We can explore the performance of ECFM in the situation where the parameterized physics is partially correct, but incomplete. We assume a source term of $s(x)=100x\sin(2\pi x)$ and take $C=10$ noiseless measurements of a numerical solution to obtain the constraints. The body force is parameterized as $b(x;\epsilon) = \epsilon \sin(2\pi x)$. See Figure \ref{pp7} for the results. The strong form recovers a physics parameter of $\epsilon=155.9$ and the weak form finds $\epsilon=155.7$. Note that it is no longer clear what value to expect for the physics parameter, as the parameterization of the missing physics is inaccurate, but not orthogonal to the true source term. The constraint force from the strong form is $z=557.2$, and the weak form gives $z=550.6$. This shows that the constraint force and the discovered source term have the same order of magnitude. Note how the constraint force is less than the previous example---this gives a quantitative measure of the improvement in the consistency between the parameterization and the true model. We claim that there is no better set of units for a quality measure than those of a force or other source term. This is because, by comparing to the magnitude of the source terms the system does or is expected to see, the analyst should be able to easily understand what a ``large'' or ``small'' force is.

\subsection{1D hyperelastic bar}

  Given that the constraint forces are explicitly controlled, the formulation of the physics loss will have no significant impact on the reconstructed solution. In this example, we use the weak form of a compressible Neohookean model of a bar to explore solution reconstruction with ECFM in the nonlinear setting. The strain energy density in the reference configuration of a 3D hyperelastic solid is given by the compressible Neohookean model \cite{bonet_nonlinear_2008} as


\begin{equation*}
    \Psi\Big( \mathbf{u}(\mathbf{X}) \Big) = \frac{\ell_1(\mathbf{X})}{2}\Big( I_1 - 3 \Big) - \ell_1(\mathbf{X}) \ln J + \frac{\ell_2(\mathbf{X})}{2} \Big( \ln J \Big)^2 ,
\end{equation*}

\noindent where $\ell_1$ and $\ell_2$ are material properties called ``Lam\'e parameters'' which may vary in space. The strain energy makes use of the following definitions:

\begin{equation*} 
    \begin{aligned}
        \mathbf{F} = \mathbf{I} + \pd{\mathbf{u}}{\mathbf{X}}, \\
        I_1 = \mathbf{F} : \mathbf{F}, \\
        J = \det(\mathbf{F}),
    \end{aligned}
\end{equation*}

\noindent where $\mathbf{u}$ is the displacement and $\mathbf{X}$ is the position in the reference configuration. In one spatial dimension, the strain energy density reduces to 

\begin{equation} \label{energy}
    \Psi\Big( u(X) \Big) = \frac{\ell_1(X)}{2}\qty( \qty(1+\pd{u}{X})^2 - 1  ) - \ell_1(X) \ln\qty(1 + \pd{u}{X}) + \frac{\ell_2(X)}{2}\qty( \ln(1+\pd{u}{X}))^2.
\end{equation}

The true solution $u(X)$ is obtained for a given source term and Lam\'e parameters by solving the weak form system, which can be derived from the condition of stationarity of the total potential energy. This reads

\begin{equation*}
    \int_0^1 \pd{\Psi}{\qty(\pd{u}{X})} \pd{\delta u}{X} - s(x)  \delta u dX = 0.
\end{equation*}

The solution and test space are discretized with $u(X) = \sum_{i=1}^N \theta_i f_i(X) $ where $f_i(X)$ is the same sine series as Eq. \eqref{fourier}. This enforces zero Dirichlet boundaries by construction. A reference solution is generated by solving this equation numerically with a standard nonlinear solver in MATLAB. The stationarity condition for the energy is also the Bubnov-Galerkin weak form of the system. The constraints are generated by sampling the solution with

\begin{equation*} 
    v_i = u(X_i) + \xi, \quad \xi \sim \mathcal{U}(-\sigma,\sigma), \quad i=1,\dots,C. 
\end{equation*}

The noise is uniformly distributed around zero. The constraint on the reconstructed solution is taken to be $v_i - \sigma \leq \hat w(x_i) \leq v_i + \sigma$. The stationarity condition of the energy with the parameterized material, body force, and explicit constraint forces is

\begin{multline} \label{hwf}
    \hat  R_j(\boldsymbol{\theta},\boldsymbol{\lambda}|\boldsymbol{\epsilon}) = \int_0^1 \ell_1(X) \qty(1+\pd{\hat w}{X}) \pd{f_j}{X}  - \ell_1(X) \frac{1}{1+\pd{\hat w}{X}} \pd{f_j}{X} + \ell_2(X)\ln(1+\pd{\hat w}{X}) \frac{1}{1+\pd{\hat w}{X}} \pd{f_j}{X} \\
    - b(x; \boldsymbol \epsilon)f_j(X) + \sum_{i=1}^C \lambda_i \Gamma(X-X_i)f_j(X) dX = 0, \quad j=1,2,\dots,N.
\end{multline}

This is the governing equation for the inner-loop problem in the ECFM, where the source term is parameterized. We have not yet discussed how to handle inequality constraints with ECFM. These can be handled with slack variables without forming the Lagrange multiplier problem shown in Eq. \eqref{L3}. Remember that if we form a Lagrange function, fictitious constraint forces are introduced to the governing equation for the physics. We can simply write an analogous system to the Karush-Kuhn-Tucker (KKT) conditions without forming the Lagrange function. This isolates the physics from constraint forces, which are beyond the analyst's control. This reads

\begin{equation} \label{bigsys}
    \mathbf{R}^{\text{ineq}}(\boldsymbol{\theta},\boldsymbol{\lambda}, \boldsymbol{s}|\boldsymbol{\epsilon}) = \begin{bmatrix} \mathbf{\hat R}(\boldsymbol{\theta},\boldsymbol{\lambda}|\boldsymbol{\epsilon}) \\
    \boldsymbol{h}(\boldsymbol{\theta}) + \mathbf{s} \odot \mathbf{s} \\ \boldsymbol{\tilde \lambda} \odot \mathbf{s}
    \end{bmatrix} = \mathbf{0},
\end{equation}

\noindent where $\odot$ is the element-wise product. See Appendix D for a derivation of this expression, in particular the dependence of $\boldsymbol{\tilde \lambda}$ on the constraint forces magnitudes $\boldsymbol \lambda$ and a discussion of the slack variables $\mathbf{s}$. The inequality constraints are rearranged to be written as $h_i(\boldsymbol{\theta}) \leq 0$ for $i=1,\dots,2C$. We have introduced slack variables $\mathbf{s}$ to enforce the inequality constraint and the ``complementary slackness'' condition \cite{ruszczynski_nonlinear_2006}. Complementary slackness requires that when the slack variable is zero, the constraint force is active, and when the slack variable is nonzero, the constraint force is inactive. The explicit constraint method problem with inequality constraints can then be written as

\begin{equation} \label{ineq}
\begin{aligned}
   & \underset{\boldsymbol{\epsilon}}{\text{argmin }} \frac{1}{2} \boldsymbol{\lambda} \cdot \mathbf{H} \boldsymbol{\lambda} \\
   & \text{s.t. } \mathbf{R}^{\text{ineq}}(\boldsymbol{\theta},\boldsymbol{\lambda},\mathbf{s}|\boldsymbol{\epsilon}) =  \mathbf{0}.
\end{aligned}
\end{equation}

  This constrained optimization problem is solved similarly to the one in the previous subsection. We note that in our attempts to compare ECFM to PINNs-type approaches, choosing numerical examples must be done carefully. In the case that the exact model is recoverable, there will be no difference in the reconstructed solution between an appropriately formulated PINNs-type penalty problem and the ECFM. In the case that the exact model is not recoverable, we do expect differences between PINNs and ECFM, except that neither will agree with the true solution, which makes comparison difficult. The ECFM promises interpretability, robustness, and data consistency, and these need not always correspond with accuracy. If the missing physics is parameterized incorrectly, there is no reason to expect to recover the true solution. The difficulty in making comparisons is exaggerated in the presence of measurement noise, which further complicates attempts to recover the exact solution. Instead, ECFM offers a predictable and customizable solution reconstruction methodology that avoids the pitfalls of PINNs. Along these lines, we will demonstrate how the predictability and customizability can be used to the analyst's advantage.

\begin{figure}[hbt!]
\centering
\includegraphics[width=0.6\textwidth]{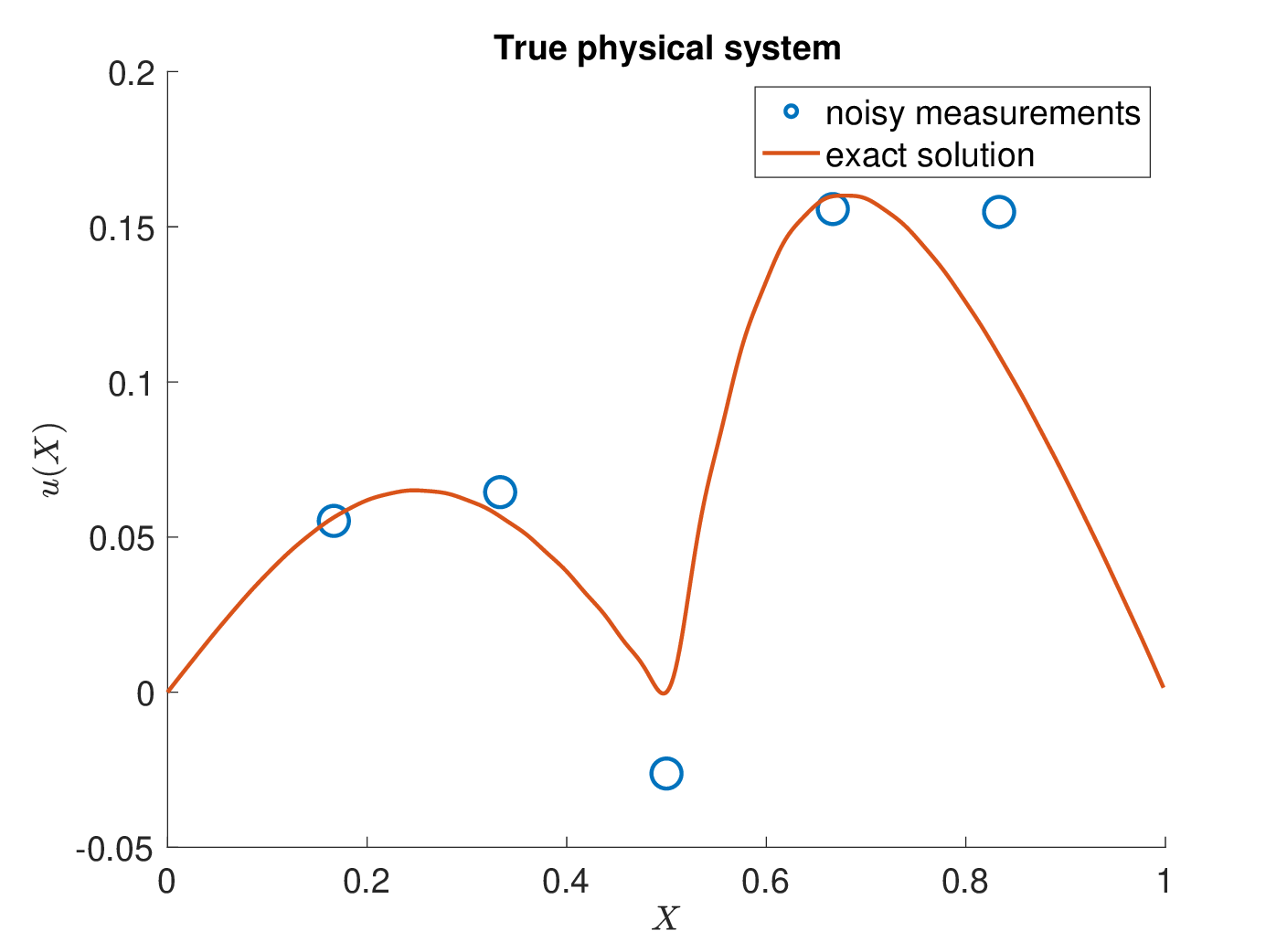}
\caption{A sample of measurement data to use in the solution reconstruction. A point force is known to be applied at $X=1/2$.}
\label{pp8}
\end{figure}

  Consider a hyperelastic bar with $C=5$ measurements with uniform noise with $\sigma=1\text{e-}2$. The reference solution is obtained with $\ell_1(X)=\ell_2(X)=1$ and $s(X)=30X$ along with a constraint of $u(1/2)=0$ enforced using Lagrange multipliers. Because we work with the variational energy, the constraint force from Lagrange multipliers is a point force, even in the nonlinear setting. The Fourier basis with $N=50$ approximates the kink in the displacement at $X=1/2$. In the context of solution reconstruction, we only have the measurement data, but we may have additional knowledge about the problem. Let us say that we know there is a point reaction force at $X=1/2$, and we seek to reconstruct the solution from measurement data like that of Figure \ref{pp8}. We parameterize the source term with $b(X;\epsilon) = \epsilon X$ and take the objective for a PINNs-based approach to be the strong form residual with inequality constraints enforced by Lagrange multipliers. Computing the strong form residual of the hyperelastic model, the optimization problem is given by 

\begin{equation} \label{hpinn}
    \begin{aligned}
    &\underset{\boldsymbol{\theta},\epsilon}{\text{argmin }} \frac{1}{2}  \int_0^1  \Bigg[ \frac{\partial^2 \hat w}{\partial X^2}\qty( 1 + 2\qty(1+\pd{\hat w}{X})^{-2} - \ln(1+\pd{\hat w}{X})\qty(1+\pd{\hat w}{X})^{-2})
    +\epsilon X \Bigg]^2 dX   \\
    &\text{s.t. } h_i(\boldsymbol{\theta}) \leq 0, \quad i=1,\dots,2C.
    \end{aligned}
\end{equation}

\begin{figure}[hbt!]
\centering
\includegraphics[width=0.98\textwidth]{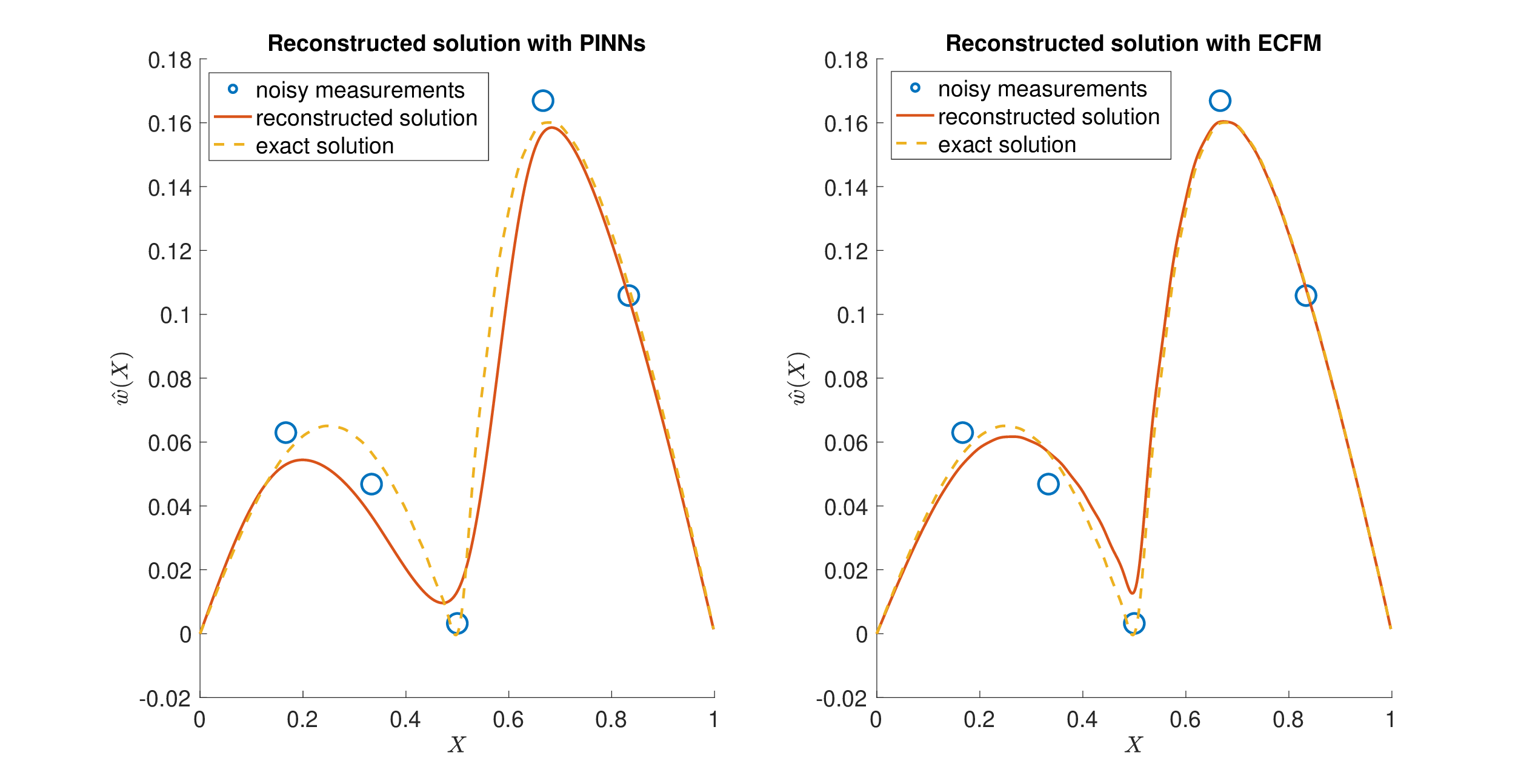}
\caption{By controlling the constraint force, the ECFM leads to a reconstructed solution which captures known features of the true solution (kink in the displacement). In this case, this also leads to a more accurate reconstruction.}
\label{pp9}
\end{figure}

The objective for ECFM is given by Eq. \eqref{ineq}. Knowing that there is a point force in the system, we will make the constraint forces

\begin{equation*}
    \Gamma(X-X_i) = \text{max}\Big(0,p(1-p|X-X_i|)\Big),
\end{equation*}

\noindent where $p$ is a parameter that controls the width of hat-shaped constraint force. As $p\rightarrow \infty$, we approximate a delta function centered at the constraint location. In practice, $p$ will be large but finite. We can now compare the reconstructed solutions from PINNs and ECFM. We take $p=20$ for the constraint force at the center of the domain to approximate a point force. See Figure \ref{pp9} for a comparison of the two reconstructed solutions. As expected, the solution reconstructed using PINNs is smoother than the exact solution with the point reaction force. This is because the constraint forces coming from Lagrange multipliers are distributed over the whole domain, just like the linear elastic problem. On the other hand, because we introduce a constraint force that approximates a concentrated reaction force, the ECFM reproduces the kink in the displacement at the center of the domain. We note that as $p$ is decreased, the two solutions should come to agree with each other. Define an error metric of the reconstructed solution as 

\begin{equation}
    \mathcal{E} = \qty( \int \Big( \hat w(\mathbf{x}) - u (\mathbf{x}) \Big)^2 d\Omega)^{1/2}.
\end{equation}

The error with the PINNs approach is $\mathcal{E}=1.2\text{e-}2$ whereas ECFM gives an error of $\mathcal{E}=4.5\text{e-}3$. We note that there is no reason to expect any method to obtain zero error in the presence of measurement noise. But as Figure \ref{pp9} shows, the ability to control the constraint force contributes to the accuracy of the reconstructed solution. This example demonstrates how the ability to control constraint forces leads to reconstructed solutions which are more predictable (in their smoothness properties, in this case) and customizable than the standard PINNs approach.

\subsection{2D heat conduction}

  Thus far, we have shown that the ECFM satisfies the criteria of interpretability, robustness, and data consistency by construction. As a final example, we will show that ECFM and PINNs produce very similar reconstructed solutions in the case that the missing physics is smooth. This helps explain why considerations about constraint forces have not surfaced in the literature, as smooth missing physics is a common scenario. This will help clarify the situations in which ECFM---an admittedly more complex formulation of the solution reconstruction problem---is called for. To this end, we will consider steady-state heat conduction with advection on the unit square. The governing PDE is given by 

\begin{equation} \label{heatconduction}
    \begin{aligned}
      &  \nabla \cdot \Big( \mathbf{A}(\mathbf{x}) \nabla u(\mathbf{x}) \Big) + \mathbf{s}(\mathbf{x}) - \mathbf{a}(\mathbf{x}) \cdot \nabla u(\mathbf{x}) = \mathbf{0}, \quad \mathbf{x} \in \Omega, \\
      &  u(\mathbf{x}) = \mathbf{0}, \quad \mathbf{x} \in \partial \Omega,
    \end{aligned}
\end{equation}

\noindent where $\mathbf{A}(\mathbf{x})$ is the conductivity matrix for the fluid medium, $\mathbf{s}(\mathbf{x})$ is a heat source, $\mathbf{a}(\mathbf{x})$ is the incompressible velocity field, and $u(\mathbf{x})$ is the temperature. We will use a multilayer perceptron neural network to discretize the temperature. The input-output relation for the $i$-th hidden layer is

\begin{equation*}
    \mathbf{y}_i = \sigma\Big(  \mathbf{W}_i\mathbf{y}_{i-1} + \mathbf{B}_i  \Big),
\end{equation*}

\noindent where $\sigma(\cdot)$ is a nonlinear ``activation function'' which is applied element-wise. As shown, the output $\mathbf{y}_i$ then becomes the next layer's input. The parameters of the neural network are the collection of the weight matrix $\mathbf{W}$ and bias vector $\mathbf{B}$ for each layer. Thus, we can write the neural network parameters as $\boldsymbol{\theta} = [ \mathbf{W}_1, \mathbf{B}_1, \mathbf{W}_2,\mathbf{B}_2,\dots]$ and the reconstructed solution as $\hat w(\mathbf{x}; \boldsymbol{\theta}$). An approximate reference solution to Eq. \eqref{heatconduction} will be obtained by discretizing the temperature with a two-hidden-layer neural network of width 10 with the Dirichlet boundary conditions built in per Eq. \eqref{dirichlet} and by minimizing the strong form loss. The problem parameters are as follows:

\begin{equation*}
\mathbf{A}( \mathbf{x})= \mathbf{I}, \quad s(\mathbf{x})=500\sin(\pi x_1) \sin(\pi x_2) e^{-100\Big( (x_1-0.25)^2 + (x_2-0.25)^2 \Big)}, \quad \mathbf{a}(\mathbf{x}) = \begin{bmatrix} 5 \\ 5 
\end{bmatrix}.
\end{equation*}

\begin{figure}[hbt!]
\centering
 \includegraphics[width=0.98\textwidth]{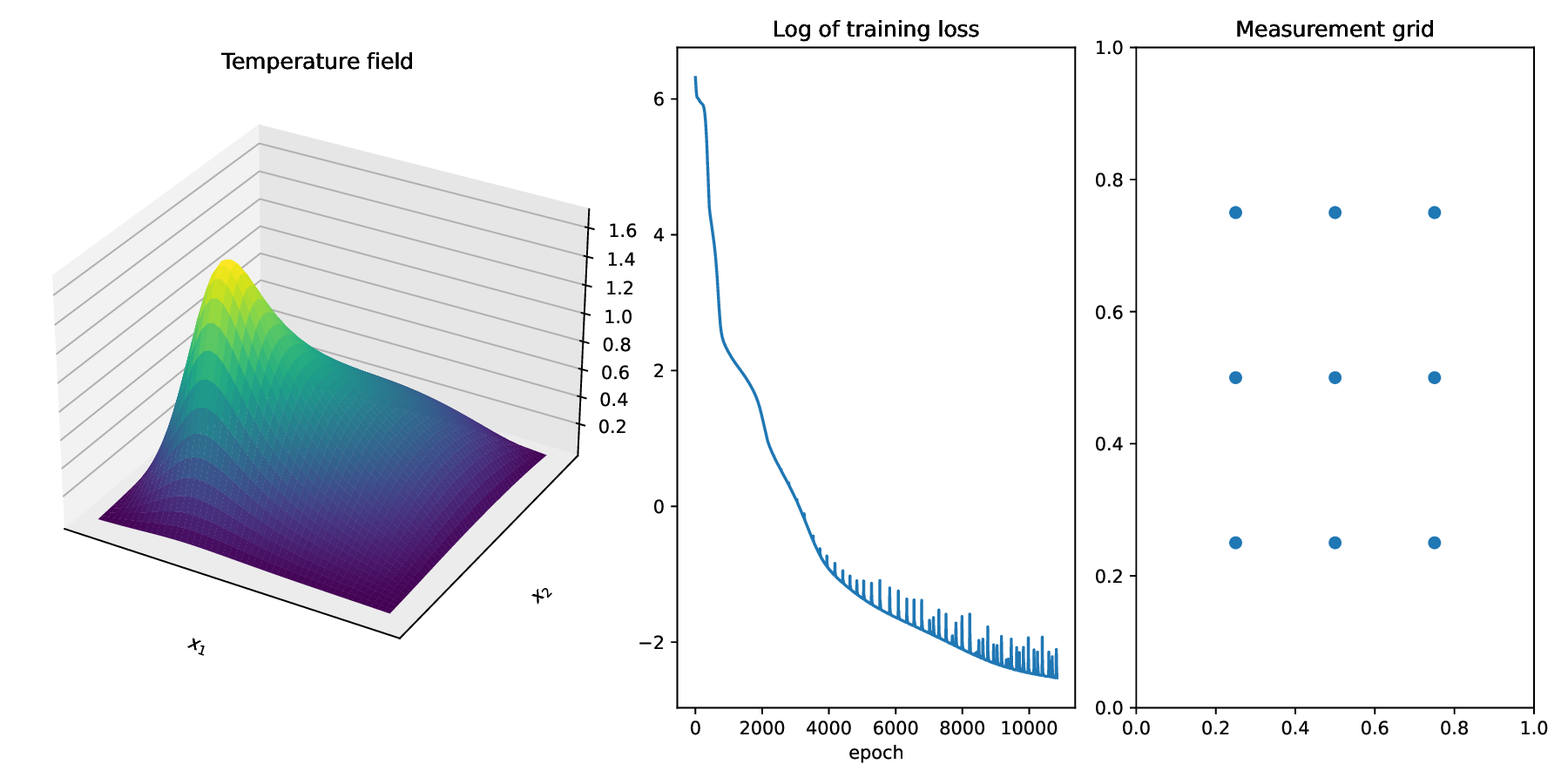}
\caption{Temperature field computed using a neural network discretization and strong form loss. The true solution is sampled noiselessly at $C=9$ positions.}
\label{pp10}
\end{figure}

We take noiseless measurements at $C=9$ points on a uniform grid for the solution reconstruction problem. The computational domain is the unit square $\Omega = [0,1] \times [0,1]$. See Figure \ref{pp10} for the reference solution, training convergence, and measurement grid. The training is cut off once the integral of the squared residual passes below a certain threshold. We assume equality constraints on the measurement data, and that the source term is known. To compare the performance of ECFM and PINNs when the physics is incorrectly parameterized, we take the conductivity to be parameterized per $\tilde A(x) = \Big(1 + \epsilon \sin(\pi x_1) \sin(\pi x_2) \Big) \mathbf{I} $ and neglect the advection term. Introducing the explicit source terms and writing out the Petrov-Galerkin weak form with test functions $f_j(\mathbf{x})$, the governing equation for the physics of the solution reconstruction problem is

\begin{equation*}
    \hat R_j( \boldsymbol{\theta}, \boldsymbol{\lambda} | \epsilon) = \int \mathbf{ \tilde A}(\mathbf{x};\epsilon)\nabla \hat w (\mathbf{x}; \boldsymbol{\theta}) \cdot \nabla f_j - s(\mathbf{x}) f_j -\sum_{i=1}^C \lambda_i \Gamma( \mathbf{x} - \mathbf{x}_i) f_j d\Omega .
\end{equation*}

The Petrov-Galerkin method appears to be preferred in the context of neural network discretization because Bubnov-Galerkin methods suffer from issues with trivial solutions. To the best of the author's knowledge, this is not explicitly stated anywhere in the literature. See Appendix E for a brief discussion of the issues with the Bubnov-Galerkin weak form. Following our previous approach, the constrained system can be written as

\begin{equation}\label{ppp}
    \mathbf{R}(\boldsymbol{\theta}, \boldsymbol{\lambda} | \epsilon) = \begin{bmatrix} \mathbf{\hat R}(\boldsymbol{\theta}, \boldsymbol{\lambda} | \epsilon) \\ \mathbf{h}(\boldsymbol{\theta})
    \end{bmatrix}.
\end{equation}

\begin{figure}[hbt!]
\centering
\includegraphics[width=0.45\textwidth]{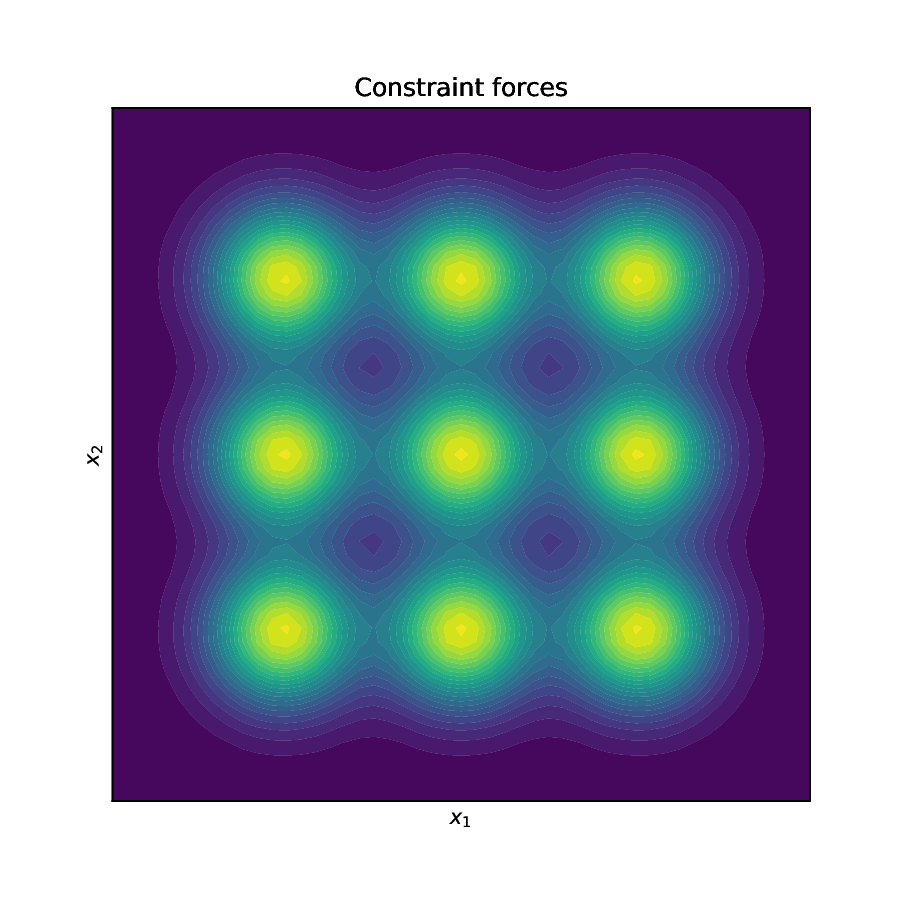}
\caption{Radial basis constraint force functions $\Gamma(\mathbf{x}-\mathbf{x}_i)$ for $C=9$ constraints. Here, the width is given by $p=25$.}
\label{cfs}
\end{figure}

In 2D, we would need to form a mesh on the computational domain in order to use finite element basis functions for the constraint forces. To avoid this, but still assume a simplistic form of constraint force, the constraint forces are taken to be radial basis functions centered at the constraint location, i.e.,

\begin{equation*} \label{rbf}
    \Gamma(\mathbf{x}-\mathbf{x}_i) = e^{-\frac{1}{2}p \lVert \mathbf{x} - \mathbf{x}_i \rVert^2},
\end{equation*}

\noindent where $p=25$ going forward. See Figure \ref{cfs} to visualize the constraint force functions. We note that the formulation of the ECFM we have used in the previous examples is best-suited for linear discretizations of the solution, where the inner-loop system $\mathbf{R}(\boldsymbol{\theta},\boldsymbol{\lambda}|\boldsymbol{\epsilon})=\mathbf{0}$ can be solved directly by inverting a block matrix for linear problems or with Newton's method for nonlinear problems. When using neural network discretizations, it is often advantageous to formulate the solution of the system as a minimization problem. This rules out Lagrange multipliers, for which it is necessary to find a saddle point \cite{ruszczynski_nonlinear_2006}. A minimization formulation can be accomplished by approximating a zero of Eq. \eqref{ppp} with

\begin{equation*} \label{stf}
\underset{\boldsymbol{\theta},\boldsymbol{\lambda}}{\text{argmin }} \frac{1}{2} \lVert \mathbf{\tilde R}(\boldsymbol{\theta}, \boldsymbol{\lambda} | \boldsymbol{\epsilon}) \rVert^2 + \lambda_d' \lVert \mathbf{h}(\boldsymbol{\theta})\rVert^2,
\end{equation*}

\noindent where $\lambda_d'$ is a new penalty parameter. In contrast to the standard PINNs objective, there is no conflict between the data penalty and satisfying the physics, regardless of the accuracy of the parameterized equation. This objective can be driven down to zero (or near zero), unlike a standard PINNs formulation, as a result of the explicit introduction of constraint forces into the system. We note that this approach is similar to \cite{zou_correcting_2024}, except that they take their constraint forces to be a deep neural network. Because there are more neural network parameters than there are data constraints, this may give rise to the data-consistency issues we discussed in Section 4. We reiterate that the expressivity of the constraint force must be matched to the number of constraints in order to prevent unnecessary introduction of non-uniqueness of the reconstructed solution. Having switched to a minimization-based inner-loop problem, it is natural to use the strong form residual instead of the residual of the weak form system. The inner-loop problem is then written as

\begin{equation}\label{ecfmobj}
\underset{\boldsymbol{\theta},\boldsymbol{\lambda}}{\text{argmin }} \frac{1}{2}  \int \Big( \nabla \cdot ( \mathbf{A}(\mathbf{x};\epsilon) \nabla \hat w) + s(\mathbf{x}) + \sum_{i=1}^C \lambda_i \Gamma( \mathbf{x}-\mathbf{x}_i)\Big)^2 d\Omega   + \lambda_d' \lVert \mathbf{h}(\boldsymbol{\theta})\rVert^2.
\end{equation}

The outer-loop problem still involves finding the physics parameter such that the constraint force is minimized---we have simply redefined how $\boldsymbol \lambda (\epsilon)$ is computed. The PINNs problem will be formulated using the strong form loss with penalties on the measurement data. This reads

\begin{equation}\label{obj}
    \underset{\boldsymbol{\theta}, \epsilon}{\text{argmin }} \frac{1}{2}\int\Big( \nabla \cdot (\mathbf{\tilde A}(\mathbf{x} ; \epsilon) \nabla \hat w )+ s(\mathbf{x}) \Big)^2 d\Omega + \frac{\lambda_d}{2}\sum_{i=1}^C\Big( \hat w(\mathbf{x}_i; \boldsymbol{\theta} ) - v_i \Big)^2.
\end{equation}

\begin{figure}[hbt!]
\centering
\includegraphics[width=1.02\textwidth]{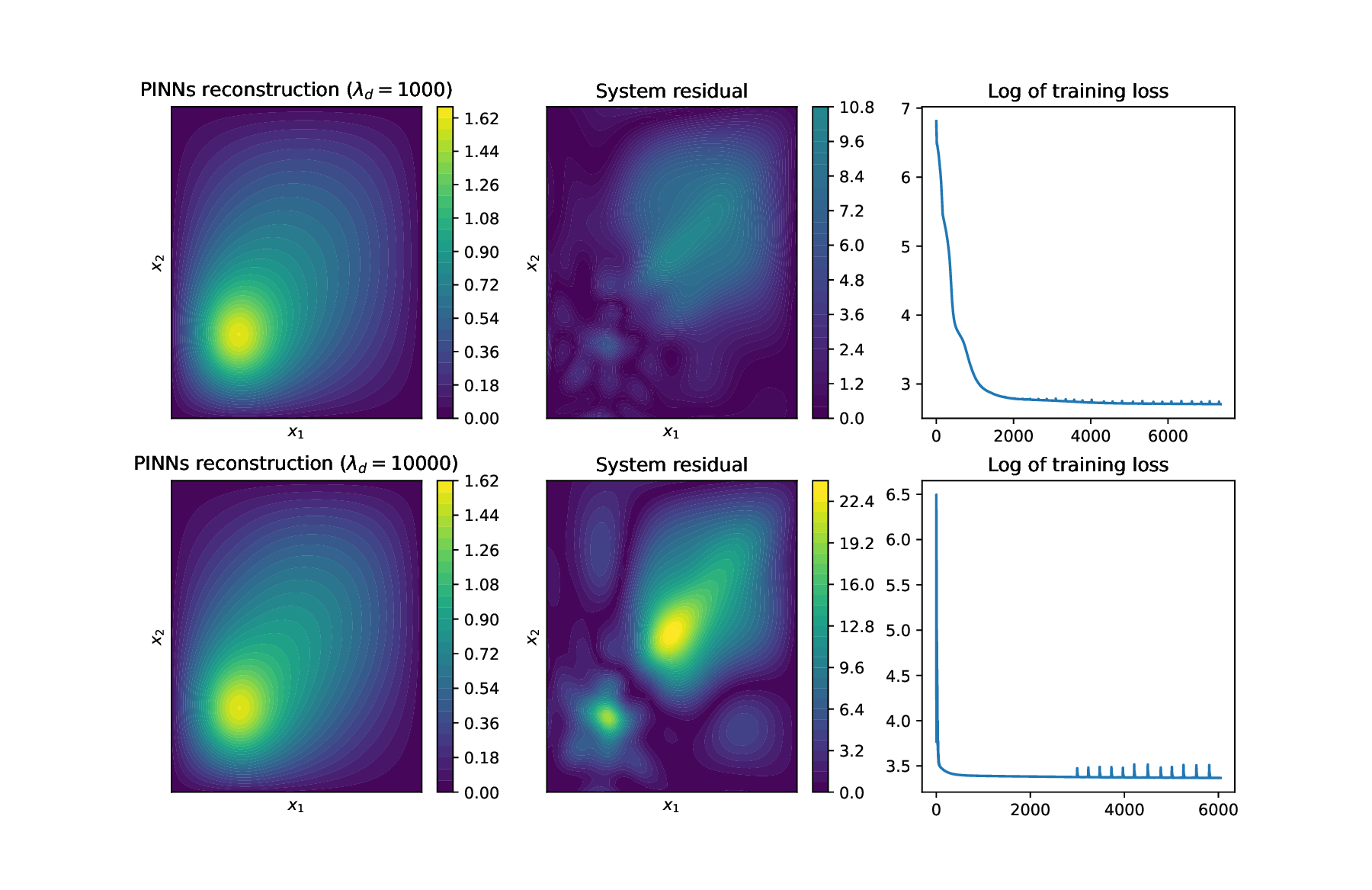}
\caption{Reconstructing the solution with the strong form loss and a penalty on the data using PINNs. The converged nonzero objective value indicates that the physics is incorrectly parameterized. The two rows show the solution obtained with two different penalty parameters.}
\label{update}
\end{figure}

Remember that $\lambda_d$ is a penalty parameter which must be chosen by the user. We can now compare the reconstructed solution from PINNs and ECFM. We will solve the problem with PINNs first. Taking $\lambda_d=1000$, we minimize the objective given by Eq. \eqref{obj} using ADAM optimization to reconstruct the solution and recover the physics parameter. We use hyperbolic tangent activation functions and two hidden layers, each with a width of 15. See Figure \ref{update} for the results from these two problems. The training is cut off once the error stagnates. We note that the objective is not zero at the minimum, indicating a conflict between satisfying the physics and satisfying the constraints. With this penalty parameter, the maximum constraint violation is $\max( |\hat w(\mathbf{x}_i) - v_i|)=0.15$. The error with the exact solution is $\mathcal{E}=9.5\text{e-}2$, and the recovered physics parameter is $\epsilon=0.67$. As seen in Figure \ref{update}, the residual of the system is computed as $\mathcal{R}=\nabla \cdot (\mathbf{\tilde A}(\mathbf{x};\epsilon) \nabla \hat w)+s(\mathbf{x})$ at the converged solution. The residual appears ``structured'' in the sense that it is concentrated in a certain region, another indication that there is missing physics. We can increase the penalty parameter to $\lambda_d=10000$ to decrease the constraint violations. The maximum constraint violation is now $\max( |\hat w(\mathbf{x}_i) - v_i|)=0.03$ which we deem satisfactory. The recovered physics parameter is $\epsilon=0.65$ and the error with the exact solution is now $\mathcal{E}=7.8\text{e-}2$, a relatively minor decrease owing to the accuracy in constraint enforcement.

\begin{figure}[hbt!]
\centering
\includegraphics[width=0.98\textwidth]{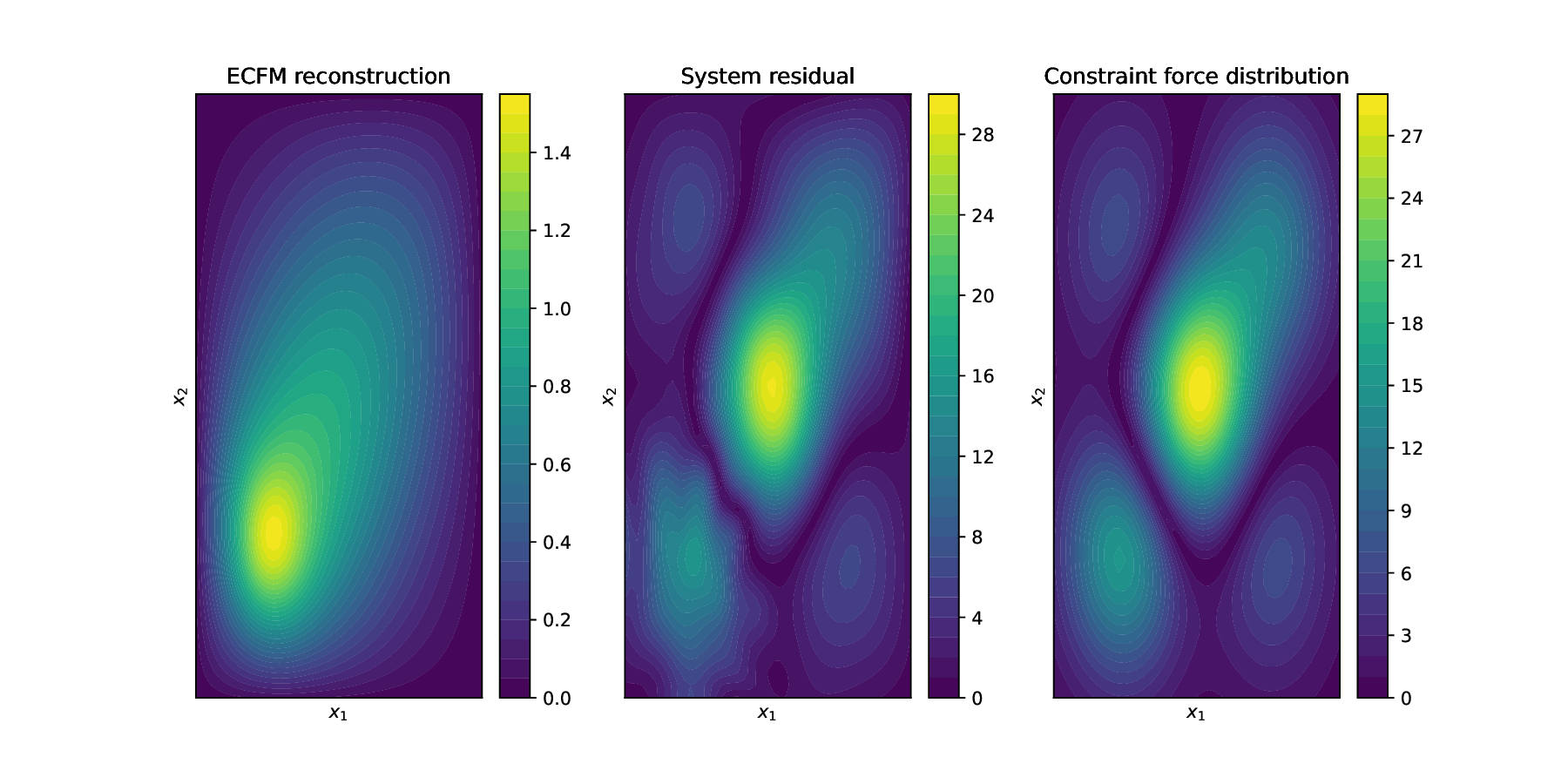}
\caption{When using ECFM, the source terms introduced to enforce the constraints are explicitly controlled. The constraint forces and the system residual are equivalent, which by virtue of being interpretable, gives the analyst clues as to the physics missing from the parameterization.}
\label{pp13}
\end{figure}

  Next, we use ECFM to minimize the constraint force with the inner-loop problem formulated per Eq. \eqref{ecfmobj} with $\lambda_d'=1000$. Note that the value of the penalty parameter is not particularly important, as both the objective and constraint violations can be driven to zero. See Figure \ref{pp13} for the reconstructed solution using ECFM. The recovered physics parameter is $\epsilon=0.63$, similar to the PINNs-based approach. The maximum constraint error violation is 0.002, with no need to select a penalty hyperparameter. The error with the exact solution is $\mathcal{E}=6.3\text{e-}2$, which is a slight improvement over the PINNs penalty method. Remember that, in general, we do not expect to recover the exact solution when the physics is inaccurately parameterized. It is also interesting to look at the difference between the reconstructed solutions using the two methods. We can compute the difference between PINNs using $\lambda_d=10000$ and ECFM as

\begin{equation*}
     \qty( \int \Big(  \hat w^{\text{PINN}} - \hat w^{\text{ECFM}}\Big)^2 d\Omega )^{1/2}= 2\text{e-}2.
\end{equation*}

The difference between the two methods is three times smaller than the difference between each method and the exact solution, thus, ECFM and PINNs are performing similarly. This can be seen from Figures \ref{update} and \ref{pp13}. In both cases, source terms are introduced into the system in order to enforce the constraints. In ECFM, the spatial distribution of the constraint forces is controlled, while they are not the case in PINNs. But, the constraint forces from each method exhibit similar smoothness properties, thus leading to similar reconstructed solutions. This contrasts with the situation where the constraint forces were localized when using the energy loss. We thus propose the following guideline: ECFM and PINNs will lead to similar reconstructed solutions when the analyst introduces smooth constraint forces. But, in spite of the similarity of the reconstructed solutions, ECFM leads to a more interpretable quality measure and is less dependent on the choice of penalty parameter. In this problem, the total constraint force $z$ is interpreted as an integrated heat flux. We have $z=37.24$ for this problem, which the analyst can use to assess the quality of the reconstructed solution. We note that our choice to measure the total constraint force as the integral of the square of the constraint source term is somewhat arbitrary---one could also take the absolute value. 

  When using PINNs, even the residual of the system cannot be interpreted purely as a missing source term. As we show for the 1D model problem in Appendix A, the penalty influences the solution operator in addition to the source term. A similar analysis can be carried out on the 2D heat conduction problem to demonstrate that the penalty term affects the discretized solution operator, complicating the interpretation of the system residual. Thus, as a final study, we can demonstrate the utility of a system residual which can be straightforwardly interpreted as a missing source term. We consider steady state heat conduction with conductivity $\mathbf{A}=\mathbf{I}$, the same source term $s(\mathbf{x})$ given above, and no missing physics parameters to fit. We use measurement data from the advection-diffusion system of Eq. \eqref{heatconduction} to reconstruct a solution, then see if we can estimate the unknown advection velocity by using the system residual. For this problem, the residuals with PINNs and ECFM are

\begin{equation}
    \mathcal{R}^{\text{PINN}}(\mathbf{x}) = - \Big( \nabla^2 \hat w(\mathbf{x}) + s(\mathbf{x}\Big),\quad \mathcal{R}^{\text{ECFM}}(\mathbf{x}) = -\sum_{i=1}^C \lambda_i \Gamma(\mathbf{x}-\mathbf{x}_i).
\end{equation}

For both methods, the advection velocity, which is assumed to be constant, is estimated with

\begin{equation}
    \underset{\mathbf{a}}{\text{argmin }} \frac{1}{2}\int \Big( \mathbf{a} \cdot \nabla \hat w(\mathbf{x}) - \mathcal{R}(\mathbf{x}) \Big)^2  d\Omega,
\end{equation}

\noindent where $\hat w(\mathbf{x})$ is the converged solution from either PINNs or ECFM and $\mathcal{R}$ is the corresponding residual. For PINNs, we obtain a solution with Eq. \eqref{obj} but with the dependence on $\epsilon$ removed. Similarly, we find the ECFM solution using Eq. \eqref{ecfmobj} with no learnable physics parameters and thus no constraint force minimization. We simply find the magnitudes of the constraint forces such that the constraints are satisfied. Note that the true advection velocity is $\mathbf{a}=[5,5]^T$. See Table \ref{tab:refinement} for the results of the recovered advection velocity components from both methods as the measurement grid is refined. As a result of the penalty method's influence on the solution operator, the residual cannot be thought of purely as a missing source term. This leads to inaccurate recovery of the advection velocity. The estimate of the advection velocity improves with ECFM as the quantity of data increases. This shows that in addition to being interpretable, the residual from ECFM gives the analyst more informative clues as to how the parameterization of the missing physics can be improved.

\begin{table}[h]
  \centering
  \begin{tabular}{|c|c|c|c|c|c|c|c|}
    \hline
    Method & $C=4$ & $C=9$ & $C=16$ & $C=25$ & $C=36$ & $C=49$ & $C=64 $\\
    \hline
    $\bar a$ (PINN) & 0.43 & 1.96 & 2.90 & 2.90 &  2.88 & 2.90 &  2.88 \\
    \hline
    $\bar a $ (ECFM) & 0.42 & 2.89 & 4.46 & 4.60 &  4.70 & 4.79 & 4.86 \\
    \hline
  \end{tabular}
  \caption{Estimating the components of the advection velocity using the PINNs and ECFM residuals for varying amounts of measurement data. In all cases, both methods predicted $a_1\approx a_2$, so we simply show their average, denoted as $\bar a $. Because the penalty term in PINNs influences the solution operator in addition to introducing a source term, the residual is not useful in fitting the missing advection term. The true advection velocity components are $a_1=a_2=5$.}
  \label{tab:refinement}
\end{table}


\section{Conclusion}

  Although our attention was restricted to static boundary value problems, we showed that the standard techniques for solution reconstruction taken in the physics-informed machine learning literature fail to reliably satisfy the three basic criteria of interpretability, robustness, and data consistency. This motivated an in-depth diagnosis of the causes of these failures with a simple model problem, which paved the way for a remedy. The concept of ``constraint forces'' proved to be the linchpin of our analysis---interpreting Lagrange multipliers as scaling source terms provided an interpretable quality measure, and it also showed why different physics losses could produce such different reconstructed solutions. The minimum constraint force principle was proposed as a unifying perspective on what constitutes optimal recovery of the model parameters. These considerations paint the picture of a rather subtle collusion between the physics loss and constraint term (whether it be from Lagrange multipliers or the penalty method), the effects of which are fictitious source terms whose magnitudes are chosen for the solution to satisfy constraints. To remedy the impacts of the unpredictable form of this constraint force and its sensitivity to the problem formulation, we introduced explicit source terms into the system and forewent standard techniques for constraint enforcement. We provided examples of how this could be accomplished. The main takeaway is as follows: anytime constraints are enforced by adding a term to an optimization objective that involves a physics loss, the constraint term will introduce fictitious forces to the system unless the objective can be driven to zero. The form of these forces will depend on the choice of the physics loss and the nature of the penalty term. These forces may or may not be appropriate for the system under study, and one needs to be careful that the minimum constraint force principle is built into the optimization problem when fitting the missing physics parameters. This led us to develop a novel strategy called the explicit constraint force method (ECFM). We have provided examples of how neglect of these considerations can lead to problems. We also note that even in the situations where PINNs and ECFM lead to very similar reconstructed solutions, we see it as an advantage to knowingly introduce constraint forces and to postulate the minimum constraint force as the notion of an optimal reconstruction from the measurement data. This brings to the surface some ideas relevant to PINNs and solution reconstruction that we believe to be insufficiently recognized.

  Some recommendations are in order for when to use PINNs and when to consider using ECFM. In the situation where parameterized physics can recover the true model, there will be no conflict between the physics loss and data penalty in PINNs, and no constraint forces will be introduced to the system. In the case that the missing physics is smooth (i.e., not point sources) but unknown, PINNs may be adequate as ECFM will produce a similar reconstructed solution. That being said, as a result of explicitly introducing source terms (and not changing the solution operator), ECFM is more interpretable. This can be a boon in trying to correct the model by studying the system residual. In the case that it is important to have a quality metric for the reconstructed solution and/or control the discretization of the missing physics, ECFM is required. 

  Future work includes extending this analysis to systems outside of solid mechanics and heat transfer. Similarly, the method could be extended to dynamical systems where the explicit constraint forces have time-dependent magnitudes. There has been little guidance given to the analyst on how to choose the spatial form of the constraint forces. An additional avenue for future research is to build parametric constraint force functions and include their parameters as additional optimization variables in minimizing the total constraint force. In the context of the 2D heat transfer problem, this could mean fitting the width of the radial basis constraint forces as well as their magnitudes to minimize the constraint force. This would help side-step potential concerns about the arbitrariness of the analyst's choice of the form of the constraint forces.


\section*{Acknowledgments}

  This work was funded by the National Defense Science and Engineering Graduate Fellowship (NDSEG) through the Department of Defense (DOD) and the Army Research Office (ARO). A. Doostan was supported by the AFOSR award FA9550-20-1-0138, with Dr. Fariba Fahroo as the program manager, and by the Department of Energy, National Nuclear Security Administration, Predictive Science Academic Alliance Program (PSAAP) Award Number DE-NA0003962.

\appendix

\counterwithin*{equation}{section}
\renewcommand\theequation{\thesection\arabic{equation}}

\section{Penalty method constraint force (model problem)}

  In this appendix, we show how to derive the constraint force corresponding to a penalty method and the strong form loss objective. The discretized form of the penalty method loss function for the model problem of Eq. \eqref{modelproblem} with a single physics parameter is 

\begin{equation*}
    \mathcal{L}_1' = \frac{1}{2}\int_0^1 \qty( \sum_j \theta_j \frac{\partial^2 f_j}{\partial x^2} + \epsilon b(x) )^2 dx + \frac{\lambda_d}{2}\Big( \sum_j \theta_j f_j(x_i) - v \Big)^2.
\end{equation*}

We assume one constraint for the sake of exposition and a discretization of the form $\hat w = \sum_j \theta_j f_j(x)$ where $f_j(x)$ are unspecified shape functions which automatically satisfy the Dirichlet boundary conditions. The governing equation for the physics is 

\begin{equation*}
    \pd{\mathcal{L}_1'}{\theta_k} = \sum_j \theta_j \qty( \int_0^1 \frac{\partial^2 f_j}{\partial x^2} \frac{\partial^2 f_k}{\partial x^2} dx ) + \epsilon \int_0^1 b(x) \frac{\partial^2 f_k}{\partial x^2} dx + \lambda_d \Big( \sum_j \theta_j f_j(x_i) - v \Big) f_k(x_i) = 0.
\end{equation*}

We can rewrite this expression using matrix-vector notation as 

\begin{equation*}
    \pd{\mathcal{L}_1'}{\boldsymbol{\theta}} = \mathbf{K} \boldsymbol{\theta} + \epsilon \mathbf{F} + \lambda_d (\mathbf{G} \cdot \boldsymbol \theta) \mathbf{G} - \lambda_d v \mathbf{G} = \mathbf{0}.
\end{equation*}

\noindent where $\mathbf{K}$ is the stiffness matrix, $\mathbf{F}$ is the force vector, and $\mathbf{G}$ is a vector of shape functions evaluated at the constraint position. The penalty method introduces two new terms to the discretized governing equation. The first term introduces a displacement-dependent force that therefore changes the operator itself. The second term is a fixed force with an unknown spatial form. The governing equation can be rewritten as

\begin{equation*}
    \Big( \mathbf{K} + \lambda_d\mathbf{G} \mathbf{G}^T \Big) \boldsymbol{\theta} + \epsilon \mathbf{F} - \lambda_d v \mathbf{G} = \mathbf{0}.
\end{equation*}

We can interpret the fixed force coming from the penalty method by making an analogy to the force vector $\mathbf{F}$. Let us define the spatial form of the constraint force as $g(x)$. We need $g(x)$ such that its corresponding force vector is the matrix $\mathbf{G}$. This occurs when

\begin{equation*}
    \int_0^1 g(x) \frac{\partial^2 f_j}{\partial x^2} dx = f_j(x_i).
\end{equation*}

Integrating by parts twice and using that the shape functions are zero on the boundaries, we have

\begin{equation*}
    \int_0^1 g(x) \frac{\partial^2 f_j}{\partial x^2} dx = \int_0^1 \frac{\partial^2 g}{\partial x^2} f_j dx + g\pd{f_j}{x} \Bigg|^1_0 = f_j(x_i).
\end{equation*}

It can be seen that this equation is satisfied when $g(x)$ is the same hat function centered at the constraint location as in Eq. \eqref{final}. Thus, the penalty method introduces distributed constraint forces but also changes the discretized operator governing the system response.

\section{Weak form constraint force (model problem)}

  In this appendix, we show how to derive the constraint force corresponding to the method of Lagrange multipliers and the weak form objective. The Bubnov-Galerkin weak form of the system is obtained by integrating the discretized governing equation against the basis functions which construct the solution approximation. This reads

\begin{equation*}
    \int_0^1 \sum_j \theta_j \frac{\partial^2 f_j}{\partial x^2} f_k + \epsilon b(x) f_k dx = -\sum_j \theta_j \int_0^1 \pd{f_j}{x} \pd{f_k}{x} dx + \epsilon \int b(x) f_k dx = 0,
\end{equation*}

\noindent where the second equality follows from integration by parts and that the shape functions are zero on the boundary. Switching to matrix-vector notation, we can write a solution to the solution reconstruction problem with a weak form system subject to a single constraint as

\begin{equation*}
\begin{aligned}
   & \underset{\boldsymbol{\theta},\boldsymbol{\epsilon}}{\text{argmin }} \frac{1}{2}\Big( \mathbf{K} \boldsymbol{\theta} - \epsilon \mathbf{F} \Big) \cdot \Big( \mathbf{K} \boldsymbol{\theta} - \epsilon \mathbf{F} \Big)  \\
   & \text{s.t. } \mathbf{G} \cdot \boldsymbol{\theta} - v = 0.
\end{aligned}
\end{equation*}

We will enforce the constraint with Lagrange multipliers. The governing equation for the physics is obtained by taking a derivative with respect to the solution parameters. This reads

\begin{equation*}
    \pd{}{\boldsymbol{\theta}}\Big[\frac{1}{2}\Big( \mathbf{K} \boldsymbol{\theta} - \epsilon \mathbf{F} \Big) \cdot \Big( \mathbf{K} \boldsymbol{\theta} - \epsilon \mathbf{F} \Big)  + \lambda( \mathbf{G} \cdot \boldsymbol{\theta} - v )  \Big] = \mathbf{K} \mathbf{K} \boldsymbol{\theta} - \epsilon \mathbf{K} \mathbf{F} + \lambda \mathbf{G} = \mathbf{0}.
\end{equation*}

We have used the symmetry of the stiffness matrix $\mathbf{K}$. Multiplying this expression by $\mathbf{K}^{-1}$, the governing equation for the physics can be written as 

\begin{equation*}
    \mathbf{K} \boldsymbol{\theta} - \epsilon \mathbf{F} + \lambda \mathbf{K}^{-1} \mathbf{G} = \mathbf{0}.
\end{equation*}

The Lagrange multiplier scales a constraint force vector given by $\mathbf{K}^{-1} \mathbf{G}$. The vector $\mathbf{G}$ is the force vector one would obtain with a source term of $\delta(x-x_i)$. Inverting the stiffness matrix against this force vector can then be thought of as a displacement response to a point force at the location of the constraint. We know that the boundary conditions are encoded by the stiffness matrix, and that this 1D linearly elastic model problem will have a piecewise linear displacement response to a point force. Thus, the form of the constraint force $\mathbf{K}^{-1}\mathbf{G}$ is yet another hat function centered at the location of the constraint.

\section{Sensitivity derivatives}

  In this appendix, we show how to compute gradients of the constraint force by treating the solution parameters $\boldsymbol \theta$ and the constraint forces $\boldsymbol \lambda$ as a function of the physics parameters $\boldsymbol \epsilon$. We can iteratively minimize the constraint force objective using gradient descent or a variant thereof. The constraint in Eq. \eqref{problem} can be enforced automatically if the constraint force is treated as a function of the physics parameters $\boldsymbol \epsilon$. In this case, the gradient of the objective is 

\begin{equation*}
    \pd{z}{\epsilon_i} = \mathbf{H} : \pd{\boldsymbol{\lambda}}{\epsilon_i},
\end{equation*}

\noindent where $i$ indexes the components of the vector of physics parameters. The requirement that the constraint force magnitudes $\boldsymbol{\lambda}$ respect Eq. \eqref{res} can be enforced automatically when computing the sensitivity derivative $\partial \boldsymbol{\lambda} /\partial \boldsymbol{\epsilon}$. Taking the gradient of Eq. \eqref{res} with respect to the physics parameters, we have 

\begin{equation*}
    \frac{d\mathbf{R}}{d\boldsymbol \epsilon} = \begin{bmatrix} \partial \mathbf{\hat R} / \partial \boldsymbol{\epsilon} \\ \mathbf{0}
    \end{bmatrix}^{\text{expl}} + \begin{bmatrix} \mathbf{\partial \hat R} / \partial \boldsymbol{\theta} & \partial \mathbf{\hat R} / \partial \boldsymbol{\lambda} \\ \partial \mathbf{\hat w} / \partial \boldsymbol \theta & \mathbf{0}
    \end{bmatrix} \begin{bmatrix} \partial \boldsymbol{\theta} / \partial \boldsymbol{\epsilon} \\ \partial \boldsymbol{\lambda} / \partial \boldsymbol{\epsilon}
    \end{bmatrix} = \mathbf{0},
\end{equation*}

\noindent where the superscript ``expl'' refers to an explicit dependence of the system on $\boldsymbol{\epsilon}$, as opposed to an implicit dependence through the solution parameters $\boldsymbol{\theta}$ and the constraint force coefficients $\boldsymbol{\lambda}$. The desired sensitivity derivative can then be computed with

\begin{equation*} \label{sens}
    \begin{bmatrix} \partial \boldsymbol{\theta} / \partial \boldsymbol{\epsilon} \\ \partial \boldsymbol{\lambda} / \partial \boldsymbol{\epsilon}
    \end{bmatrix} = \begin{bmatrix} \partial \mathbf{\hat R} / \partial \boldsymbol{\theta} & \partial \mathbf{\hat R} / \partial \boldsymbol{\lambda} \\ \partial \mathbf{\hat w} / \partial \boldsymbol{\theta} & \mathbf{0}
    \end{bmatrix}^{-1} \begin{bmatrix} -\partial \mathbf{\hat R} / \partial \boldsymbol{\epsilon} \\ \mathbf{0}
    \end{bmatrix}^{\text{expl}}.
\end{equation*}

This ensures that the constraints in Eq. \eqref{problem} are respected automatically during the optimization. Notice that the constraints are enforced without a penalty method or Lagrange multipliers. This ensures that the constraints are handled by forces (source terms) that are within the analyst's control.

\section{Inequality constraints with explicit constraint force method}

  We require a technique to enforce inequality constraints on the reconstructed solution without introducing new constraint forces as source terms. This rules out forming the Lagrange function with the residual of the system and Lagrange multipliers enforcing the displacement constraints. We can try to write out the analogue of the KKT conditions without passing through the Lagrange function. The problem is that we have $2C$ constraints and only $C$ constraint forces. The number of constraints doubles in the inequality setting because there are upper and lower bounds on the displacement at each constraint location. The remedy to the mismatch between the number of constraints and constraint forces, which prevents us from enforcing the analogue of the complementary slackness condition, is to introduce a separate constraint force for the upper and lower bounds of the constraint. The weak form residual system is then

\begin{multline} \label{aug}
    \tilde R_j(\boldsymbol{\theta},\boldsymbol{\lambda}|\boldsymbol{\epsilon}) = \int_0^1 \ell_1(X) \qty(1+\pd{\hat w}{X}) \pd{f_j}{X}  - \tilde \ell_1(X) \frac{1}{1+\pd{\hat w}{X}} \pd{f_j}{X} + \tilde \ell_2(X;\epsilon_1)\ln(1+\pd{\hat w}{X}) \frac{1}{1+\pd{\hat w}{X}} \pd{f_j}{X} \\
    -\epsilon_2 b(x)f_j(X) + \sum_{i=1}^C \Big( \lambda^L_i \Gamma(X-X_i) - \lambda^U_i \Gamma(X-X_i)\Big)f_j(X) dX = 0, \quad j=1,2,\dots,N
\end{multline}

\noindent where $\boldsymbol{\tilde \lambda} = [\boldsymbol{\lambda}^U, \boldsymbol{\lambda}^L]^T$ is the collection of the constraint forces corresponding to the upper and lower limits of the inequality constraints on the displacement. The minus sign scaling $\boldsymbol{\lambda}^U$ is important due to the ``dual feasibility'' condition from Lagrange multipliers, which typically says that $\boldsymbol{\lambda} \geq 0$. In our problem, this will read $\boldsymbol{\tilde \lambda} \geq 0$. The dual feasibility condition is a way of eliminating spurious solutions. If any component of $\boldsymbol{\tilde \lambda} $ is less than zero, this means that the constraint force is activated to force the displacement from the feasible region to satisfy the constraint as an equality. This is undesirable in finding the minimum constraint force such that the solution is feasible. We have doubled the number of constraint forces to match the number of constraints, but note that for each $i=1,\dots,C$, only one of the two $\lambda_i^L, \lambda_i^U$ will be non-zero. This means that we have the following relation between the original constraint forces and the augmented set:

\begin{equation}
    \lambda_i = \begin{cases} \lambda_i^L & \text{if }\lambda_i^L \neq 0 \\
    -\lambda_i^U & \text{if } \lambda_i^U \neq 0
    \end{cases}.
\end{equation}

This is a consequence of the complementary slackness condition. We can say abstractly that $\boldsymbol{\lambda} = \boldsymbol{\lambda}(\boldsymbol{\tilde \lambda})$. This means that the introduction of additional constraint forces does not change the physics of the problem---it is simply a means to handle the inequality constraints. The analogue to the KKT conditions for the explicit constraint force method with inequality constraints is then

\begin{equation}
    \mathbf{R}^{\text{ineq}}(\boldsymbol{\theta},\boldsymbol{\lambda}, \boldsymbol{s}|\boldsymbol{\epsilon}) = \begin{bmatrix} \mathbf{\tilde R}(\boldsymbol{\theta},\boldsymbol{ \lambda}|\boldsymbol{\epsilon}) \\
    \boldsymbol{h}(\boldsymbol{\theta}) + \mathbf{s} \odot \mathbf{s} \\ \boldsymbol{\tilde \lambda} \odot \mathbf{s}
    \end{bmatrix} = \mathbf{0}.
\end{equation}

\noindent where $\odot$ is the element-wise product and $\{ s_i \}_{i=1}^{2C}$ are the slack variables corresponding to the constraints and augmented set of constraint forces. Note that we have not enforced the positivity of the constraint forces yet. This can be done by simply taking the absolute value of the constraint forces in Eq. \eqref{aug}, which enforces the positivity of the constraint forces automatically. Because these equations do not come from an expanded optimization problem in the form of a Lagrangian, it is possible to make modifications of this sort without needing to adjust the other equations in the system.

\section{Trivial solutions with Bubnov-Galerkin weak form}

  In this appendix, we show that using the weak form to find solutions to PDEs discretized by neural networks presents problems. Consider a solution to the model problem given in Eq. \eqref{model} using a Bubnov-Galerkin weak form with the displacement discretized as a nonlinear function of the parameters $\boldsymbol{\theta}$. This is the case when we employ a PINNs solution, as the solution depends nonlinearly on the weights and biases of the neural network. One way to obtain the governing equation is to discretize the energy and then take its gradient with respect to the parameters. A more illustrative route to a solution is to apply the principle of Galerkin orthogonality to the nonlinear discretization. Galerkin orthogonality says that a solution to a discretized differential equation is obtained when the residual is orthogonal to the space in which the solution is approximated. In the case of a nonlinear discretization, Galerkin orthogonality states that the residual is orthogonal to the local tangent of the manifold on which the solution lives. When the solution $u(x)$ is approximated by $\hat u(x;\boldsymbol{\theta})$, the tangent vector to the approximation space is given by $\partial u / \partial \boldsymbol{\theta}$. In this case, the weak form of the model problem with the nonlinear discretization is given by

\begin{equation} \label{exa}
    \mathbf{R}(\boldsymbol{\theta}) = \int \frac{\partial^2 \hat u}{\partial x^2} \pd{\hat u}{\boldsymbol{\theta}} - s(x) \pd{\hat u}{\boldsymbol{\theta}} dx = \mathbf{0}.
\end{equation}

\begin{figure}[hbt!]
\centering
\includegraphics[width=0.7\textwidth]{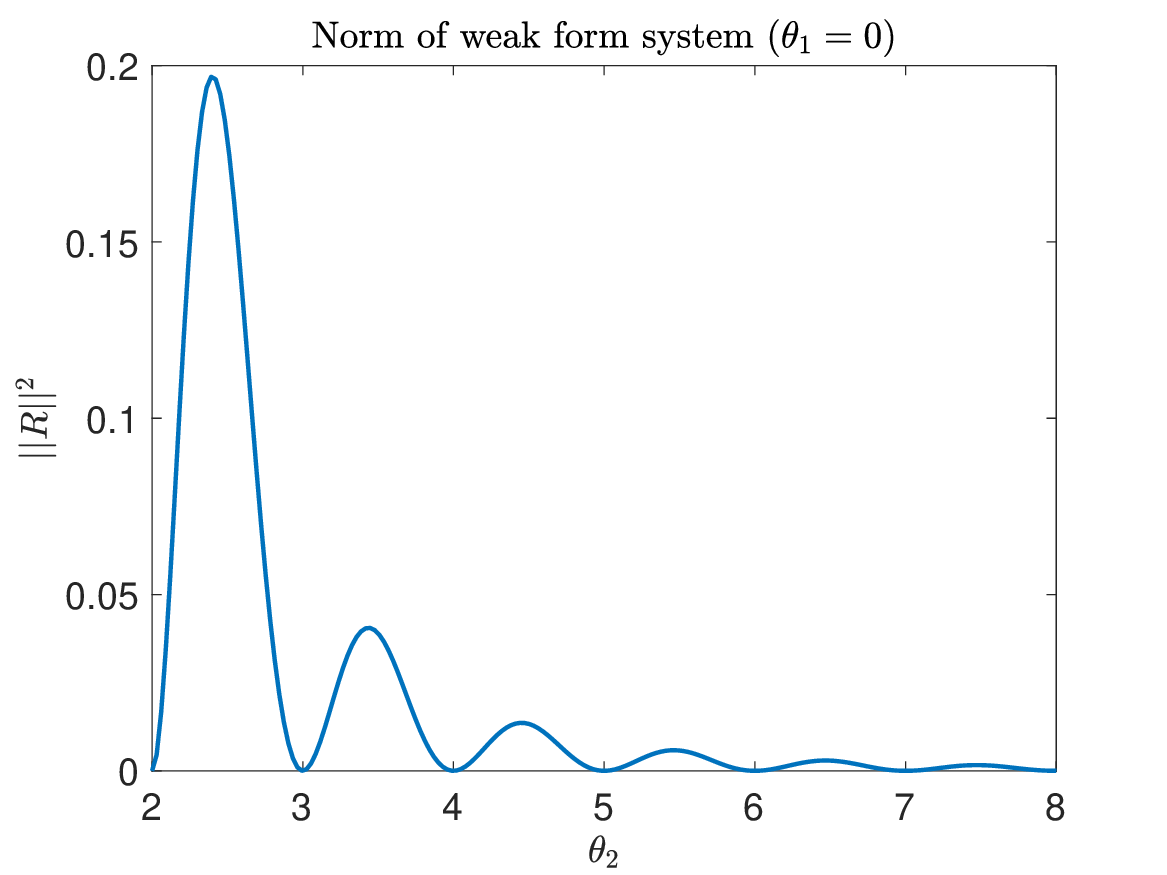}
\caption{There is an infinite family of trivial solutions to the weak form system given by Eq. \eqref{exa}. This is even though the discretization can represent the exact solution.}
\label{append}
\end{figure}

Consider a simple problem. The source term is given by $s(x) = \pi^2 \sin(\pi x)$, which corresponds to a solution of $u(x) = \sin(\pi x)$. We approximate the solution as $\hat u(x) = \theta_1 \sin(\theta_2 x)$. The exact solution is recoverable with this approximation. We can form the residual system given in Eq. \eqref{exa} and solve it using a nonlinear solver. The solver sometimes returns the exact solution $\boldsymbol{\theta}=[1,1]^T$, and other times it gives $\boldsymbol{\theta}=[0,2]^T$. As Figure \ref{append} shows, there are an infinite family of solutions $\boldsymbol{\theta}=[0,n]^T$ for $n=1,2,\dots$. In analogy with more traditional discretizations, we can think of the parameters that contribute to the solution nonlinearly as forming the ``basis'' of the approximation, and the parameters with linear influence as the coefficients on the basis. Remember that the weak form system can be satisfied for any choice of basis. Thus, with the nonlinear discretization, the system of equations in Eq. \eqref{exa} can be satisfied by ``learning'' a poor choice of basis that the trivial solution $\hat u=0$ is the best approximation. This is what happens in this example. When $\theta_2$ is an integer other than 1, the true solution is orthogonal to the basis, and the trivial solution $\theta_1=0$ satisfies the weak form system. There is no clear way to avoid trivial solutions of this sort in the process of solving the residual system. These trivial solutions occur in the case of neural network discretizations as well, where it appears in some cases easier for the network to learn collinear basis elements that are orthogonal to the true solution than to work with a more expressive basis. This issue with trivial solutions is avoided by enforcing that the error is orthogonal to a fixed function space with the Petrov-Galerkin method. This appears to be the reason why Bubnov-Galerkin methods have been avoided in the literature thus far. Finding ways to avoid these trivial solutions when using neural network discretizations is potentially an interesting area of further research.

\bibliographystyle{plain}
\bibliography{My_Library}  

\end{document}